\documentclass{elsart}
\usepackage{graphicx,amssymb,amsmath,times,lineno}
\journal{NIM}
\begin{document}
\begin{frontmatter}
\title{Artificial Neural Network based gamma-hadron segregation methodology for TACTIC telescope.}
\author{V.K.Dhar\corauthref{cor}},
\corauth[cor]{Corresponding author.}
\ead{veer@barc.gov.in}
\author{A.K.Tickoo}, \author{M.K.Koul}, \author{R.Koul}, \author{B.P.Dubey$^\dagger$}, \author{R.C.Rannot}, \author{K.K.Yadav}, \author{P.Chandra}, \author{M.Kothari}, \author{K.Chanchalani}, \author{K.Venugopal}.
\address{Bhabha Atomic Research Centre,\\
         Astrophysical Sciences Division.\\
        $^\dagger$ Electronics and Instrumentation Services Division,\\
         Mumbai - 400 085, India.}
     
\begin{abstract}
The sensitivity of a Cherenkov imaging telescope is strongly dependent on the rejection of the cosmic-ray background events. The methods which have been used to achieve the segregation  between the gamma-rays from the source and the background cosmic-rays, include  methods like Supercuts/Dynamic Supercuts, Maximum likelihood classifier, Kernel methods, Fractals, Wavelets and random forest. While the  segregation potential of the neural network classifier has been investigated in the past with modest results,  the  main purpose of this paper is to study the gamma / hadron segregation potential of various  ANN algorithms, some of which are supposed to be  more powerful in terms of better convergence and lower error compared to the commonly used Backpropagation  algorithm. 
The results obtained  suggest that Levenberg-Marquardt method outperforms all other methods in the ANN domain. Applying  this ANN algorithm  to  $\sim$ 101.44 h   of  Crab Nebula  data collected  by  the TACTIC telescope, during Nov. 10, 2005 - Jan. 30, 2006,  yields   an excess  of $\sim$ (1141$\pm$106) with a statistical  significance of $\sim$ 11.07$\sigma$, as against   an  excess  of  $\sim$ (928$\pm$100) with  a statistical significance of $\sim$ 9.40$\sigma$  obtained  with   Dynamic Supercuts selection  methodology.  The main  advantage  accruing  from the  ANN methodology is that it is more effective at higher energies and this has allowed us to re-determine the Crab Nebula energy spectrum in the energy range $\sim$ 1-24 TeV.   
\end{abstract}
\end{frontmatter}

\section{Introduction}
\label{1}
Gamma-ray photons in the TeV energy range ( 0.1-50 TeV ), to which we shall confine our attention here, are expected to come from a wide variety of cosmic objects within and outside our galaxy. Studying this radiation in detail can yield valuable and quite often, unique information about the unusual astrophysical environment characterizing these sources, as also on the intervening intergalactic space [1-3]. While this promise of the cosmic TeV $\gamma$-ray probe has been appreciated for quite long, it was the landmark development
of the imaging technique and the principle of stereoscopic imaging, proposed by Whipple [4] and the HEGRA [5] groups, respectively, that revolutionized the field of ground-based very high-energy (VHE)  $\gamma$-ray astronomy. 
\par
The success of VHE $\gamma$-ray astronomy, however depends critically on the efficiency of $\gamma$/hadron classification methods employed. Thus, in order to improve the sensitivity of ground based telescopes, the main challenge is to improve the existing $\gamma$/hadron segregation methods to efficiently reduce the background cosmic  ray   contamination and at  the same  time also retain higher number  of  $\gamma$-ray events. Detailed Monte-Carlo simulations, pioneered by Hillas [6], show that the differences between Cherenkov light emission from air showers initiated by $\gamma$-rays and protons (and other cosmic-ray nuclei) are quite pronounced, with the proton image being broader and longer as compared to the $\gamma$-ray image.
This  led to the development and successful usage of several image parameters in tandem, a technique referred to as the Supercuts/Dynamic Supercuts method. Although the efficiency of this $\gamma$/hadron event classification methodology, has been confirmed by the detection of several $\gamma$-ray sources by various independent groups including us,  there  is a need to search for still more sensitive/efficient algorithms for $\gamma$/hadron segregation. The conventionally used  Supercuts/Dynamic Supercuts method, though using several image parameters simultaneously, with some of them also being energy dependent, is still a one dimensional technique, in the sense that the parameters it uses for classification are treated separately and the possible correlations among the parameters are  ignored. 
\par
The  multivariate analysis  methods,  proposed  by various  groups, for  discriminating  between   $\gamma$-rays and hadrons are the  following:  Multidimensional Analysis based on Bayes Decision Rules [7],  Mahalonobis Distance [8],  Maximum Likelihood [9], Singular Value Decomposition [10], Fractals and Wavelets [11,12] and Neural Networks [13,14]. The comparative performance of different  multivariate classification methods like Regression ( or Classification)  trees, kernel methods, support vector machines, composite probabilities, linear discriminant analysis  and  Artificial  Neural  Networks  (ANN) has  also  been  studied  by  using Monte Carlo simulated data for  the  MAGIC telescope.  A detailed compilation of this study is reported in [15]. The  results  published  in  the above  work  indicate  that while as  the performance 
of Classification Trees, Kernel and  Nearest-Neighbour  methods  are very close to each other, the different ANN method employed (feed-forward, random search and multilayer perceptron)  yield  results  over a wide range. The feed-forward method gives a significance of $\sim$ 8.75 $\sigma$, whileas multilayer perceptron gives a somewhat poorer significance of $\sim$ 7.22$\sigma$ [15].  The discrimination  methods  like Linear Discriminant Analysis and Support Vector Machines  are  found  to be  inferior compared to others [15]. The  authors  of  the above work  claim that the  Random Forest method  outperforms  the classical methodologies.  
\par     
Details  regarding  implementation of the Random Forest method for the MAGIC   telescope and  some of  the  other  recent $\gamma$/ hadron separation methods  developed  by the H.E.S.S and VERITAS collaboration  can  be found  in  [16-20]
\par
The paper is organized  in  the following manner. Section 2 will cover a summary of some applications  where  ANN  has been used. Salient  design features  of the TACTIC  telescope and   generation of  simulated  data  bases will be presented  in sections 3 and 4, respectively. Section 5  covers  the definition  and statistical analysis  of various  image parameters. A short introduction  to ANN  methodology  and a brief  description  of the ANN algorithms  used  in the present  work  have  been  presented in section 6 so that the manuscript can be followed  by researchers  who are not experts in the field  of  neural networks. Application of the ANN based  $\gamma$/hadron  methodology to TACTIC telescope will be presented in  sections 7 and 8. These two  sections cover the details  about training, testing, validation  and comparison  of various  ANN algorithms used in the present  work. Application of   the  ANN methodology  to the Crab Nebula and Mrk 421 data collected with the TACTIC telescope is presented  in sections 9. A comparison between the Dynamic Supercuts and ANN analysis methods is described in section 10 and in  section 11 we present our conclusions. 

\section{Brief  description  of  some  applications  where  ANN  have been  used}

Research activity in the last decade or so has established that ANN based algorithms are  promising alternatives to many conventional classification methods. The advantages of ANN over the conventionally used methods are mainly the following: Firstly,  ANN are data driven, self- adaptive methods, since they adjust themselves to given data without any  explicit specification of the functional form for the underlying model. Secondly, they are universal function approximators as they can approximate any function with an arbitrary accuracy [21]. Third and most important,  ANN are able to estimate the posterior probability which provides the basis of establishing classification rule and performing statistical analysis [22,23] .  These statistical methods, though important for classification are merely based on bayesian decision theory in which posterior probability plays a central role. The fact that ANN can provide an estimate of posterior probability implicitly establishes the strong connection between the ANN and statistical methods. A direct comparison between the two, however, is not possible as ANN are non-linear and model free methods, while as statistical methods are mostly linear and model based. 
\par
Artificial neural networks have been applied quite extensively to particle physics experiments including separating gluon from Quark jet [24] and identification of the decays of the Z$^\circ$ boson into $b\bar{b}$ pairs [25].  
Application of feed-forward ANN classifier, employed by the DELPHI collaboration, for separating hadronic decays of the Z$^\circ$ into c and b quark pairs has resulted in an improved determination with respect to the standard analysis [26]. Superior performance of the Neural Network  approach, compared  to   other   multivariate  analysis  methods  including  discriminant  analysis  and  classification trees,   has   been  
reported  by   LEP/SLC  [27],   for  tagging   of  Z$^\circ$ $\longrightarrow$ $b\bar{b}$ events. Details related to application of ANN to general astronomical applications can be found in [28].
\par 
Several $\gamma$-ray astronomy groups have already explored the feasibility of using ANN for $\gamma$/ hadron separation work. While nobody  has so far worked with primary ANN ( i.e  using  Cherenkov images itself as inputs to ANN), the results reported are mainly from the use of secondary ANN   where  various image parameters  are used as inputs to the ANN. 
In an  attempt to examine the potential  of  ANN   for   improving    the efficiency  of  the imaging  technique,    $\gamma$-ray   and  proton  acceptance  of  $\sim$ 40   $\%$  and $\sim$ 0.7 $\%$, respectively was  achieved  by  Vaze  [29]   by  using   8  image  parameters  as inputs  to the  ANN. A  detailed study of applying ANN to imaging telescope data was attempted by  Reynolds and Fegan [14]   and   results  of their  study  indicate  that the ANN  method   although   being   superior  to   other  methods  like  maximum  likelihood and singular value decomposition   does   not  yield  better  results  than the  Supercuts  Method. 
The  work  reported   by  Chilingarian in [13]  by  using   8  image  parameters  as inputs  to  the  ANN,  on the other hand, indicates  a  slightly    better   performance  of  the ANN  method  as compared  to  the   Supercuts  procedure.  Using  a network  configuration of  4:5:1  on the  Whipple  1988-89  Crab Nebula  data,  the  author has  reported only   marginal  enhancement  in  the   statistical  significance  ( viz.,   $\sim$35.80$\sigma$  as against      $\sim$34.30$\sigma$   obtained  with   the Supercuts  method), but   there  is  a significant  increase  in the   number   $\gamma$-rays   retained  by the  ANN ( viz.,  $\sim$3420   as against  $\sim$2686   obtained  with   the  Supercuts  method).  
Application  of   Fourier   transform  to  Cherenkov  images   and   then  using   the   resulting    spatial frequency 
components  as  inputs  to a Kohonen unsupervised neural network   for   classification  has   been reported   by Lang 
[30].  The  performance  of   Multifractal and Wavelet parameters was examined by the  HEGRA collaboration  in [31] by  using a data sample  from the  Mkn 501  observation.  The  authors  of  the  above  work  report  that  combining  Hillas and multifractal parameters  using a neural network  yields  a slight improvement in performance as compared to the Hillas parameters  used alone. 
\par
There are also many other assorted [32,33] and non-imaging applications including  data collected by extensive  air shower arrays  where  ANN have  been applied.   Bussino and Mari  [34] employed a backpropagation based ANN model for separating electromagnetic and hadronic showers detected by an air shower array. They achieved a $\sim$ 75 $\%$  identification for $\gamma$-rays  and $\sim$ 74$\%$  identification  for protons.  Maneva et al [35] used a ANN algorithm for the CELESTE data.  Dumora et al [36] have also reported promising results for CELESTE data where ANN method was used for discriminating  the $\gamma$ /hadron Cherenkov events for the wavefront sampling  telescope.  The standard Sttutgart Neural Network Simulator (SNNS) package  has also  been used   for $\gamma$/hadron segregation  for the data obtained from AGRO-YBJ experiment [37].  Application  of backpropagation based ANN method   for  separating  $\gamma$/hadron   events    recorded  by   the   HEGRA air  shower array  has  been  studied  by Westerhoff et al [38].
\par
Keeping in view  the  encouraging  results   reported  in  the above  cited  literature, in particular   the   results published in  [13, 15],  we studied the $\gamma$/ hadron segregation potential of various  ANN algorithms, by applying them to the Monte Carlo simulated data. The  idea of applying  ANN for  determining  the  energy of the  $\gamma$-rays,   from  a point  source,   has   already been used  by us  [39] for  determining   the  energy spectra  of   the  Crab Nebula,  Mrk421  and Mrk501,  as measured  by the TACTIC telescope.

\section{TACTIC Telescope}
The TACTIC (TeV Atmospheric Cherenkov Telescope with Imaging Camera) $\gamma$-ray telescope has been in operation at Mt. Abu ( $24.6^o N$, $72.7^oE$, 1300m asl), a hill resort in Western India, for the last several years for the study of TeV gamma ray emissions from celestial sources.  The telescope deploys a  349-pixel  imaging camera, with a uniform pixel size   of $\sim$ $0.31^o$ and a $\sim$ $5.9^o$$\times$$5.9^o$ field-of-view,   to record  atmospheric Cherenkov events produced by an incoming cosmic-ray particle or a $\gamma$-ray photon.  The  TACTIC light-collector  uses 34   front-face aluminum-coated,  glass spherical mirrors  of 60 cm diameter  each  with  a focal length $\sim$ 400cm. The  point-spread  function  has a  HWHM of $\sim$ 0.185$^0$ ($\equiv$12.5mm)  and  D$_{90}$  $\sim$ 0.34$^0$ ($\equiv$22.8mm). Here, D$_{90}$ is   defined as the diameter of a circle, concentric  with the centroid of the image, within which 90$\%$  of reflected rays lie.  The innermost 121 pixels (11 $\times$ 11 matrix) are used for generating the event trigger, based on a predecided trigger criterion which  is   either    Nearest Neighbour Pairs (NNP)  or  Nearest Neighbour  Non-collinear Triplets.  Apart from  generating  the  prompt trigger with  a coincidence  gate width  of  $\sim$18ns, the trigger generator has  a provision for producing a chance coincidence output based on  $^{12}$$C_{2}$  combinations from various groups of closely spaced 12 channels.

The data acquisition and control system of the telescope [40] is designed around a network of PCs running the QNX (version 4.25) real-time operating system. The triggered events are digitized by CAMAC based 12-bit Charge to Digital Converters (CDC) which have a full scale range of 600 pC. The relative gain of the photomultiplier tubes is monitored regularly  once in 15 minutes by flashing a blue LED, placed at a distance of $\sim1.5m$ from the camera. The data acquisition and control of the TACTIC is handled by a network of PCs. While one PC is used to monitor the scaler rates and control the high voltage of the photomultipliers (PMT), the other PC handles the data acquisition of the atmospheric Cherenkov events and LED calibration data. These two front-end PCs, referred to as the rate stabilization and the data acquisition nodes respectively, along with a master node form the multinode Data Acquisition and Control network of the TACTIC Imaging telescope. The telescope has a pointing and tracking accuracy of better than  $\pm$3 arc-minutes. The tracking accuracy is checked on a regular basis with so called "point runs", where an optical star having its declination close to that of the candidate $\gamma$-ray source is tracked continuously for about 5 hours. The point run calibration data (corrected zenith and azimuth angle of the telescope when the star image is centered) are then incorporated in the telescope drive system software  or  analysis software   so that appropriate corrections can be applied either directly in real time  or   in  an   offline   manner   during    data analysis.  
\par
The telescope records a cosmic-ray event rate of $\sim$ 2.0 Hz at a typical zenith angle of $15^o$ and is operating at a $\gamma$-ray threshold energy of $\sim$ 1.2 TeV. The telescope has a 5$\sigma$ sensitivity of detecting the Crab Nebula in 25 hours of observation time and has so far detected $\gamma$-ray emission from the Crab Nebula, Mrk 421 and Mrk 501. Details of the instrumentation aspects of the telescope, results obtained on various candidate $\gamma$-ray sources, including the energy  spectra obtained from Crab Nebula, Mrk 421 and Mrk 501,  are discussed in  [41-47].
\section{Simulation methodology and data-base generation}

We have used the CORSIKA (version 5.6211) air shower simulation code [48], with the Cherenkov option, for generating the simulated data-base for $\gamma$-ray and hadron showers. This data-base is valid for Mt. Abu observatory altitude of 1300m with appropriate values of 35.86 $\mu$T and 26.6 $\mu$T, respectively for the horizontal and the vertical components of the terrestrial magnetic field. The first part of simulation work comprised generating the air showers induced by different primaries and recording the relevant raw Cherenkov data. Folding in the light collector characteristics and PMT detector response was performed in the second part. We have generated a simulated data-base of $\sim$ 39000 $\gamma$-ray showers in the energy range 0.2-27 TeV with an impact parameter   up to 250 m. These showers are generated at 5 different zenith angles ($\theta$ = 5$^0$, 15$^0$, 25$^0$, 35$^0$ and 45$^0$). Similarly a data-base of about $\sim$ 40000 proton initiated showers, in the energy range 0.4-54 TeV, within the field of view of $\sim$ $6.6^0 \times 6.6^0$ around the pointing direction of the telescope, has also been generated by us. It  is  important  to mention  here  that the  number  of  gamma-ray showers   as  well  as the  number  of  proton  showers have  not been  generated  according  to a   power law  distribution.  However,  appropriate  $\gamma$-ray and  proton spectra,  with  differential  spectral indices   of  $\sim$ -2.6 and  $\sim$ -2.7, respectively  have been used while preparing the relevant data files used in the present work.  Wavelength dependence of atmospheric absorption, spectral response of the PMT's, reflection coefficient of mirror facets and light cones used in the imaging camera have also been taken into account while generating the data.  The obscuration encountered due to the telescope mechanical structure by the incident and reflected photons, during their propagation, is also considered. The Cherenkov photon data-base, consisting of the number of photoelectrons registered by each pixel is then subjected to noise injection, trigger condition check and image cleaning. The resulting two dimensional 'clean' Cherenkov image of each triggered event is then used to determine the image parameters for shower characterization.  Details  of simulation  aspects of the telescope  and  some  of the  results  obtained  like  effective collection area,  differential and  integral  trigger  rates  are discussed in [49].
\section{Definition  and  statistical analysis  of Cherenkov image parameters}

\subsection{Definition of Cherenkov image parameters}
A Cherenkov imaging telescope records the arrival direction of the individual Cherenkov photons and the appearance of the recorded image depends upon a number of factors like the nature and the energy of the incident particle, the arrival direction and the impact point of the particle trajectory on the ground. The principle of detecting $\gamma$-rays through the imaging technique is depicted in Fig.1a and Fig. 1b.  Segregating the very high-energy $\gamma$-ray events from their cosmic-ray counterpart  is achieved  by  exploiting  the subtle differences that exist in the two dimensional Cherenkov image characteristics (shape, size and orientation) of the two event species.  Gamma-ray events give  rise  to shower images which  are preferentially oriented towards the source position in the image plane. Apart from being narrow and compact  in shape, these images  have a cometary shape with their  light distribution  skewed  towards  their source position  in the image plane  and  become more elongated as the impact parameter increases.  On the other hand, hadronic  events give rise  to images that are, on average, broader and longer and are randomly oriented  within the field of view of the camera.
\begin{figure}[h]\centering
\includegraphics*[width=1.0\textwidth,angle=0,clip]{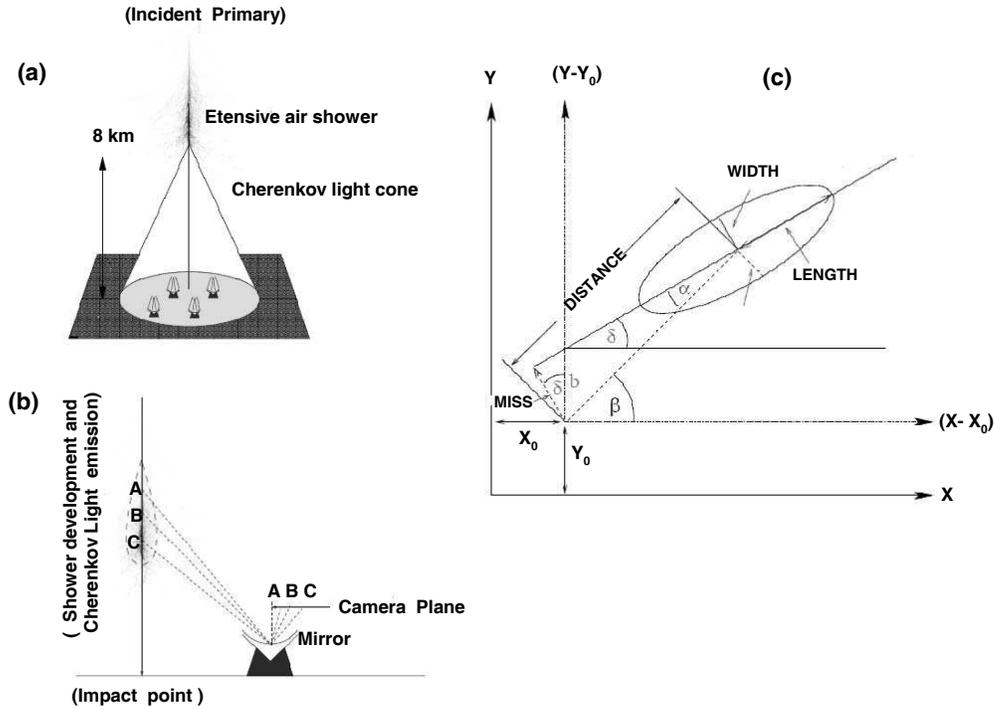}
\caption{\label{fig ---}  (a) The principle of detecting gamma-rays through the imaging technique  (b) Formation of Cherenkov  
image  in the focal plane. (c) Definition  of   Hillas  parameters characterizing   each image and  used for  rejecting  the cosmic-ray background. The  ellipse  represents  the  approximate  outline of the shower image in the focal plane of the telescope.} 
\end{figure}
For each image, which is  essentially  elliptical in shape, Hillas  parameters [6, 50] are calculated  to characterize its shape and orientation.  The  parameters,  as depicted in Fig.1c,  are obtained  using  moment analysis and  are defined as :  LENGTH--  The rms spread of light along the major axis of the image (a measure of the vertical development of the shower); WIDTH --  The rms spread of light along the minor axis of the image (a measure of the lateral development of the shower); 
DISTANCE-- The distance from the centroid of the image to the centre of the field of view; 
($\alpha$)--The angle between the major axis of the image and a line joining the centroid of the image to the position of the 
source in the  focal plane; SIZE --  Sum of all  the signals recorded  in the  clean  Cherenkov image; FRAC2-- The degree  of light concentration as determined from the ratio of the  two largest PMT signals to sum of all signals ( also referred to as Conc.). In the pioneering work of the Whipple Observatory [4], only  one parameter (AZWIDTH) was  used   in selecting   $\gamma$-ray  events. Later, the technique  was refined   to  Supercuts / Dynamic Supercuts  procedure    where  cuts based on the WIDTH and  LENGTH  of  the image as well as its orientation  are used for  segregating  the gamma rays from   the background cosmic-rays [50] 
\par
\subsection{ Statistical analysis  of  various parameters  for  selecting  the optimal features}
The success of any classification technique depends on the proper selection of the variables which are to be used for the event segregation and  the  agreement  between  the expected and the  actual  distributions  of   these variables.   Fig.2  shows the distributions of the image parameters  LENGTH, WIDTH, DISTANCE and  $\alpha$  for simulated protons and for the actual Cherenkov events  recorded by the  telescope. 
\begin{figure}[h]
\centering
\includegraphics*[width=1.0\textwidth,angle=0,clip]{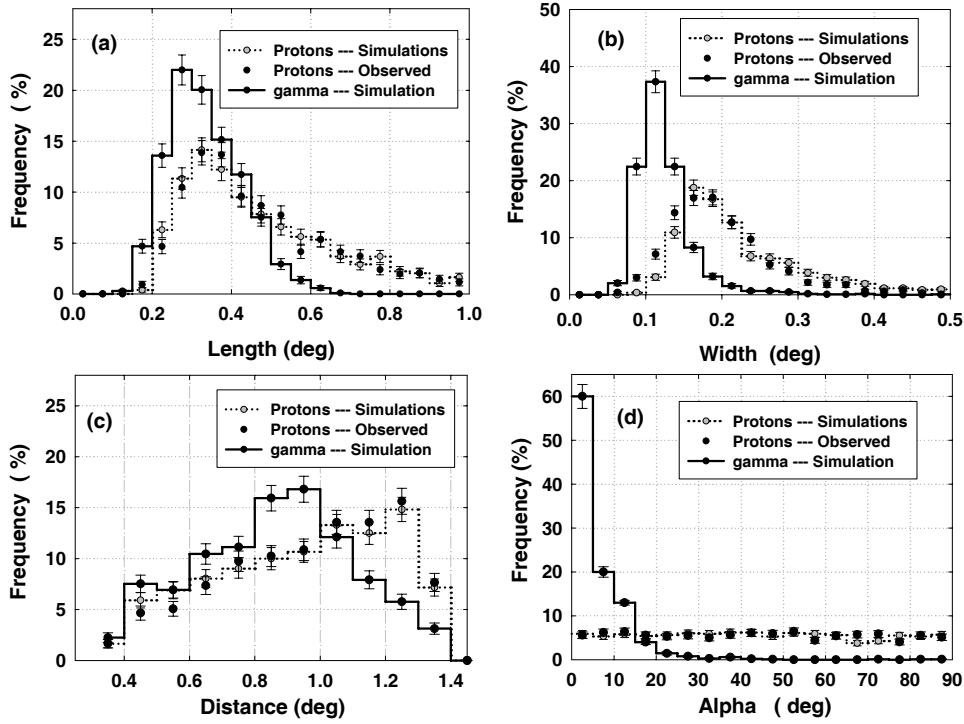}
\caption{\label{fig ---} Comparison of image parameter distributions  (a) LENGTH, (b) WIDTH, (c) DISTANCE and 
(d) ($\alpha$)  from real and the Monte Carlo simulated data for proton events. The simulated  image parameter distribution of  $\gamma$-rays  has also been shown in the figure for comparison.} 
\end{figure} 
The data plotted  here  has been  first  subjected  to  pre-filtering  cuts with  $SIZE\geq$ 50 photoelectrons (pe) and (0.4$^0 \leq DISTANCE \leq 1.4^0$) in order to ensure  that the  events  recorded  are  robust  and well contained in the camera.   
The simulated  image parameter distribution of  $\gamma$-rays  has also been shown in the figure for comparison.  The observed image parameter distributions are found to closely match the distributions  obtained from simulations for proton-initiated  showers, thus   suggesting   that  the response of the telescope  is  reasonably close  to that  predicted by simulations. For  converting the event SIZE,  recorded in charge to digital counts,  to  corresponding  number of photoelectrons,  we  have used a conversion factor of  1pe $\cong$6.5 counts [42].  In order  to understand  and  improve upon  the existing $\gamma$/hadron segregation methods  it  is   important  to   estimate  the   discriminating  capability  of  each of the Cherenkov image parameters and  their correlations[7].  The   image parameters  considered for this correlation study are :  SIZE, LENGTH, WIDTH, DISTANCE, FRAC2 and $\alpha$.
\par
In order  to   select  image parameters  which  are  best suited for  $\gamma$/ hadron separation   we have  applied  the  following  tests:
Student's t-test, Welch's t-test, Mann – Whitney U-test ( also known as Wilcoxon rank-sum test) and  
the Kolmogorov - Smirnov test (K–S test) [51].  The  Student's t-test and Welch's t-test belong to the category of parametric  tests which assume that the data are sampled from populations that follow a Gaussian distribution. While as, the Students unpaired t- test assumes that the two populations have the same variances, the Welch's t-test is a  modification of the t- test which does not  assume equal variances. Tests that do not make any assumptions about the population distribution are referred to as nonparametric tests. Mann – Whitney U-test and  Kolmogorov - Smirnov test belong to this category of tests. While the nonparametric tests are appealing because they make fewer assumptions about the distribution of the data, they are less powerful than the parametric tests. This means that the corresponding probability values tend to be higher, making it harder  to detect real differences as being statistically significant. When large data samples are considered, the difference in power is minor. Furthermore, it is worth mentioning here that the parametric tests are robust to deviations from Gaussian distributions, so long as the samples are large.
\par
In order to  apply  the above mentioned tests  to  simulated data of $\gamma$-ray  and  proton  initiated showers  we have used  $\sim$ 6000  events each,  at a zenith angle of  25$^0$ and  the  results of   these one-dimensional tests  are  summarized in Table 1.\\
\begin{table}[h]
\begin{center}
\caption{ Statistic values  of  various parametric and non-parametric statistical tests.  
Larger value  of the statistic  indicate   that  corresponding  probability  of  rejecting  the null hypothesis,   that the 
$\gamma$-ray  data   sample  and   the proton-data  sample  come  from the same population, is low.}
\begin{tabular}{|c|c|c|c|c|}
\hline
$			    $   &   $ Student's $   &   $ Welch's   $ &   $ Mann-Whitney  $  &   $  KS      $\\ 
$				  $   &   $  t-test   $   &   $  t-test   $ &   $  U-test       $  &   $  D-test       $\\
\hline
$				  $   &   $    t      $   &   $  t        $ &   $  z            $  &   $   D      $\\
\hline
$ SIZE    $   &   $    1.95   $   &   $    1.94   $ &   $ 8.66          $  &   $  0.09  $ \\
\hline
$ LENGTH  $   &   $  138.80   $   &   $    138.75 $ &   $ 90.20         $	&    $	0.85 $	\\
\hline
$ WIDTH   $   &   $  120.96   $   &   $   120.28  $ &   $ 84.75         $	&    $	0.76 $	\\
\hline
$ DISTANCE$   &   $   19.65   $   &   $   19.64   $ &   $ 17.18         $	&   $	0.18 	$	\\
\hline
$ FRAC2   $   &   $   200.84  $   &   $   200.94  $ &   $ 92.69         $	&   $	0.90 $	\\
\hline
$ ALPHA   $   &   $   112.57  $   &   $   112.53  $ &   $82.89          $	&   $	0.76 $	\\
\hline
\end{tabular}
\end{center}
\end{table}
\\
\\
Since  the P-values (i.e  the probability  of  rejecting  the null hypothesis that the $\gamma$-ray data sample and the proton-data sample come  from the same population) are usually  very  small we have instead  used the  value of the corresponding statistic for rejecting or accepting  the  null hypothesis. In other words t-statistic  values are given  in the Table. 1 for expressing the results of Student's t-test and Welch's t-test. Similarly, for Mann – Whitney U test the  z-statistic values are  given in the table (where $z=(U-m_U)/\sigma_U$  with  $m_U$ and  $\sigma_U$ as  the   mean and the standard deviation of U). For   the   Kolmogorov – Smirnov test   we have calculated D-statistic  (i.e  maximum vertical distance  between  the  two cumulative   frequency distributions). On examining  Table 1 it is  evident  that  four image parameters  (viz.,  LENGTH, WIDTH,  FRAC2 and $\alpha$) have a significant potential of providing efficient  $\gamma$/ hadron separation. Larger the value of the corresponding  statistic, lower is  corresponding  probability  of  rejecting  the null hypothesis that the $\gamma$-ray  data sample  and  the proton-data  sample  come  from the same population. 
\par
In order to estimate the statistical relationship between two image parameters for $\gamma$-ray data sample and the proton-data samples separately  we have also calculated  the Pearson product-moment correlation coefficient. Following the standard procedure,  it is obtained by dividing the covariance of the two variables by the product of their standard deviations. The closer the coefficient is to either -1 or 1, the stronger the correlation between the variables. The  results of this study, obtained separately for $\gamma$-ray and proton-data samples, are presented in  Tables  2 and 3, respectively. 
\\
\begin{table}[h]
\begin{center}
\caption{Correlation matrix for simulated $\gamma$-ray data sample at a zenith angle of 25$^0$. The values listed below for each correlation coefficient (numbers within parentheses) are the corresponding z-statistic values obtained using Fisher transformation.} Notations used are SIZ=SIZE, LEN=LENGTH, WID=WIDTH, DIS=DISTANCE, FR2=FRAC2. 
\begin{tabular}{|c|c|c|c|c|c|c|}
\hline
$			   $  &   $  SIZ   $ &  $ LEN $  & $ WID   $ & $DIS $ & $  FR2  $ & $ \alpha$\\ 
\hline
$ SIZ   $  &   $1.000    $ &  $ 0.394   $ & $ 0.474   $ & $0.072    $ & $-0.441   $ & $-0.037 $\\ 
$				 $  &   $(---)   $ &  $(33.206) $ & $(41.692) $ & $(5.603)  $ & $(38.051)$ & $(2.881)$\\
\hline
$ LEN $  &   $0.394    $ &  $ 1.000   $ & $ 0.615   $ & $0.038    $ & $-0.709   $ & $0.196 $\\
$				 $  &   $(33.206) $ &  $ (---)     $ & $(60.452) $ & $(2.908)  $ & $(78.069)$ & $(15.466)$\\
\hline
$ WID  $  &   $0.474    $ &  $ 0.615   $ & $ 1.000   $ & $-0.396   $ & $-0.569   $ & $0.456 $\\
$				 $  &   $(41.692) $ &  $(60.452) $ & $(---)      $ & $(33.360)$ & $(53.680)$ & $(39.649)$\\
\hline
$ DIS$ &   $0.072    $ &  $ 0.038   $ & $-0.396   $ & $1.000    $ & $-0.034   $ & $-0.366 $\\
$				 $  &   $(5.603)  $ &  $(2.908)  $ & $(-33.360)$ & $(---)      $ & $(-2.615) $ & $(30.491)  $\\
\hline
$ FR2  $  &   $-0.441   $ &  $ -0.709  $ & $-0.569   $ & $-0.034   $ & $1.000    $ & $-0.064 $\\
$				 $  &   $(38.051) $ &  $(78.069)$ & $(53.680)$ & $(-2.61)    $ & $(---)   $ & $(4.927)$\\
\hline
$ \alpha  $  &   $-0.037   $ &  $  0.196  $ & $ 0.456   $ & $-0.366   $ & $-0.064   $ & $1.000 $\\
$				 $  &   $(2.881)  $ &  $(15.466) $ & $(39.649) $ & $(30.491) $ & $(4.927) $ & $ (---) $\\
\hline
\end{tabular}
\end{center}
\end{table}
\begin{table}[h]
\begin{center}
\caption{ Correlation matrix   for  simulated  proton  data sample at a zenith angle of  25$^0$.  The values listed below for each correlation coefficient 
(numbers  within  parentheses)  are  the corresponding z-statistic values  (obtained  using   Fisher transformation). }
\begin{tabular}{|c|c|c|c|c|c|c|}
\hline
$			   $  &   $ SIZ   $   &   $ LEN  $ &  $ WID $ & $DIS $   &  $  FR2  $  & $\alpha$\\ 
\hline
$ SIZ   $  &   $ 1.000  $   &   $ 0.036   $ & $ 0.273   $  & $-0.301  $  &  $-0.083   $  & $-0.008$\\
$				 $  &   $ (--)    $   &   $(2.757)  $ & $(21.950) $  & $(2.332)$  &  $(6.472) $  & $(0.624) $\\
\hline
$ LEN $  &   $ 0.036  $   &   $1.000    $ & $ 0.360   $	& $-0.023  $  &  $-0.618   $  & $ 0.086$\\
$				 $  &   $(2.757) $   &   $ (-- )    $ & $(29.916) $  & $(1.792)$  &  $(60.947)$  & $(6.691) $\\
\hline
$ WID  $  &   $0.273  $   &   $0.360    $ & $1.000    $	& $ -0.036 $  &  $ -0.510  $  & $0.005 $\\
$				 $  &   $(21.950)$   &   $(29.916) $ & $( -- )    $  & $(2.806)$  &  $(46.998)$  & $ (0.368) $\\
\hline
$ DIS$ &   $-0.301  $   &   $-0.023   $ & $ -0.036  $	& $ 1.000  $  &  $ -0.006  $  & $0.012 $\\
$				 $  &   $(2.332)$   &   $(1.792) $ & $(2.806) $  & $ ( -- )  $  &  $(0.481) $  & $ (0.958) $\\
\hline
$ FR2  $  &   $-0.083  $   &   $-0.618   $ & $ -0.510  $	& $-0.006  $  &  $ 1.000   $  & $-0.028$\\
$				 $  &   $(6.472)$   &   $(60.947)$ & $(45.998)$  & $(0.481)$   &  $ ( -- )   $  & $ (2.154)$\\
\hline
$ \alpha $  &   $-0.008  $   &   $0.086    $ & $0.005   $	& $0.012   $  &  $ -0.028  $  & $1.000 $\\
$				 $  &   $(0.624)$   &   $(6.691)  $ & $(0.368)  $  & $(0.958) $  &  $(2.154) $  & $( -- )  $\\
\hline
\end{tabular}
\end{center}
\end{table}


The values of  the  t-statistic corresponding  to  each  correlation coefficient  are  also  given in these Tables (numbers  within  parentheses). These values  can  be  used  for assessing  the  significance of  the correlation.  Larger value  of the z-statistic  indicates  that  the  corresponding  probability  of  rejecting  the null hypothesis that the  observed value  comes  from a population in which  correlation coefficient $\sim$ 0,  is low. If the correlation coefficient  is   $\rho$  the  Fisher transformation can be defined as:
\par
\begin{equation}
z=\frac{1}{2}ln\left(\frac{1+\rho}{1-\rho}\right) 
\end{equation}
The Fisher $\rho$-to-z transformation [52] has also been   applied to assess the significance of the difference between two correlation coefficients ( say  $\rho_1$ and  $\rho_2$)  found in two independent samples.  The  relevant  expression  to  calculate 
this  is  given by :
\par
\begin{equation}
z_{12}=\left( \frac{|\rho_1 - \rho_2|}{\sqrt{\frac{1}{n_1-3} + \frac{1}{n_2-3}}}\right) 
\end{equation}
\par
where $\rho_1$ and $\rho_2$ are the  two  correlation coefficients, $n_1$ and $n_2$ are respectively the number of data points 
used  while  calculating   $\rho_1$ and $\rho_2$.
Table 4 gives the values for the Fisher matrix of  various  image  parameters  for the simulated  $\gamma$/proton sample.
\\
\begin{table}[h]
\begin{center}
\caption{ Fisher Matrix  for the  simulated  $\gamma$/hadron data sample  at a zenith angle of  25$^0$.   The matrix  can be  used  to assess the significance of the difference between two correlation coefficients.}
\begin{tabular}{|c|c|c|c|c|c|c|}
\hline
$			    $  &   $ SIZ  $   &   $ LEN$ &  $ WID $  & $DIS $ & $ FR2 $  & $\alpha$\\ 
\hline
$ SIZ    $  &   $ --   $   &   $ 20.83 $ &  $ 12.87 $  & $    5.6  $ & $21.30 $  & $1.59$\\
\hline
$ LEN  $  &   $ 20.83 $   &   $  --   $ &  $ 18.65 $	 & $    3.32 $ & $8.95  $  & $ 6.16$\\
\hline
$ WID   $  &   $ 12.87 $   &   $18.65  $ &  $ --  $	 & $ 20.97  $ & $ 4.56 $  & $26.70$\\
\hline
$ DIS$  &   $ 5.60  $   &   $ 3.32  $ &  $ 20.97$	 & $  --     $ & $1.51 $  & $21.68$\\
\hline
$ FR2   $  &   $21.30 $   &   $8.95  $ &  $ 4.56 $	 & $   1.51 $ & $  --   $  & $1.96$\\
\hline
$ \alpha $  &   $1.59  $   &   $6.16   $ &  $ 26.7  $	 & $ 21.68 $ & $ 1.96 $  & $ -- $\\
\hline
\end{tabular}
\end{center}
\end{table}
\\
\par
On examining  Tables 2, 3 and 4, one can select the image parameters  for  achieving  optimum $\gamma$/hadron  segregation.
This can be done  on the basis of identifying parameters  for  which  the difference between  their  correlation  coefficients is maximum.  As seen in   Table 4,   WIDTH-$\alpha$  pair  yields  the largest Fisher test value. Furthermore, it is also encouraging to find   that  the other  well known  characteristics of Cherenkov image parameters are in good agreement with our  results. For example,   dependence of the  image shape  parameters (i.e LENGTH and WIDTH)  on SIZE  for  $\gamma$-rays.  Both these parameters  yield  positive correlation coefficient of $\sim$0.394 and $\sim$0.474 as shown in Table 2. Since SIZE parameter of an image  provides  an approximate  estimate of the $\gamma$-ray  primary energy both  these parameters are expected  to be  correlated with  the event SIZE.  The  modification of the Supercuts procedure to Dynamic (or extended) Supercuts   follows  the same  principle. Negative correlation  between  DISTANCE and $\alpha$  for  $\gamma$-rays  coming  from a point source is also seen in Table 2  in accordance   with the  expected  relationship  between  these image parameters. Thus, on the basis  of results  presented in Tables 2, 3 and 4, one can confidently  say that there is a  sufficient  scope  for  utilizing  the  differences in the correlation  between various image parameters for developing  alternate  $\gamma$/hadron  segregation  methodologies. 
\par
Keeping  in view   the  fact  that,  for proton initiated  showers (as also in general  for  other cosmic-ray primaries),  the image parameter $\alpha$  is  expected  to be  independent  of other image parameters  because of the  isotropic nature  of  the cosmic-rays  we   will not use  it in the  ANN-based $\gamma$/hadron  segregation  methodology.  Justification  for  following  this  approach  is  also  evident  in  Table 3,  where   for   the   proton  data  sample,  one   finds   negligible  correlation   between  $\alpha$  and other image parameters. Thus, for extracting   the $\gamma$-ray  signal   from the cosmic-ray background, we will  use the frequency  distribution  of the $\alpha$ parameter  for the  ANN selected events.
The distribution  is expected  to  be flat  for cosmic-rays  and  should   reveal  a peak  at smaller  $\alpha$  values  for
$\gamma$-rays  coming  from a point source.  In all, we  will  use  the following   six  image  parameters  in the ANN-based    $\gamma$/hadron  segregation  methodology :  Zenith angle ($\theta$), SIZE, LENGTH, WIDTH, DISTANCE  and  FRAC2. Use of $\theta$ angle as an additional variable can be justified by keeping in view the fact that as $\theta$ angle increases,  the  line of sight  distance  to the shower maximum  also  increases, making  all projected  dimensions  of the shower (i.e, LENGTH and WIDTH) smaller. The shape  parameters  LENGTH and WIDTH  are  expected   to  approximately  scale as  $\propto$ $cos(\theta)$.  

\section{ANN  methodology   and a brief  description of algorithms  used}

A neural network is a parallel distributed information processing structure consisting of processing elements (which can process a local memory and carry out localized information processing operations) interconnected together with unidirectional signal channels called connections. Each processing element has a single output connection which branches into many collateral connections as desired. All of the processing that goes on within each processing element must be completely local , i.,e it must depend upon only the current values of the input signals arriving at the processing element via impinging connections and upon the values stored in local memory of the processing elements . ANNs like humans, learn by example, and can be configured for a specific problem through a learning process that involves adjustments of the synaptic connections, called weights which exist between neurons. A network is composed of a number of interconnected units, each unit having an input/output characteristics. The output of any unit is determined by its I/O characteristics, its interconnection to other units and external inputs. The feed-forward ANN is the simplest configuration and is constructed using layers where all nodes in a given layer are connected to all nodes in a subsequent layer. The network requires at least two layers, an input layer and an output layer. In addition to this, the network can include any number of hidden layers with any number of hidden nodes in each layer. The signal from the input vector propagates through the network layer by layer till the output layer is reached. The output vector represents the predicted output of the ANN and has a node for each variable that is being predicted. 
\par
Depending upon the architecture in which the individual neurons are connected and the error minimization scheme adopted, there can be several possible ANN configurations. While algorithms like Standard backpropagation (along with its varients like the backprop-momentum, Vanilla backprop,  Quickprop) and the Resilient backpropation come under the category of Local search algorithms, Conjugate Gradient methods, Levenberg-Marquardt algorithm and One Step Secant belong to the category of Global search algorithm. Hybrid  algorithm category constitutes models like Higher Order Neuron and Neuro Fuzzy systems. The Standard  Backpropagation network [53],  is the most thoroughly investigated  ANN algorithm till date. Backpropagation using gradient descent however converges very slowly. 
The success of this algorithm in solving large-scale problems, although depends critically  on user-specified learning rate and momentum parameters, there are however no standard guidelines for choosing these parameters. The Resilient backpropagation(RProp) algorithm was proposed by Reidmiller [54], to expedite the learning of a backpropagation algorithm. Unlike the standard Backpropagation algorithm, RProp uses only partial derivative signs to adjust weight coefficients. In the above backprop based gradient descent algorithms, it is difficult to obtain a unique set of optimal parameters, due to the existence of multiple local minima. The presence of these local minima, hampers the search for global minimum because these algorithms frequently get trapped in local minima regions and hence, incorrectly identify a local minimum as the global minimum.
\par
 The conjugate scale gradient algorithms [55]  initially use the gradient to compute a search direction and then a line search algorithm is used, to find the optimal step size along a line in the search direction. The Levenberg
algorithm involves the use of "blending method" between the steepest descent method employed by the backpropagation/resilient algorithm and the quadratic rule employed in conjugate algorithms. The original Levenberg algorithm was improved further by Marquardt, resulting in the Lavenberg-Marquardt algorithm [56]  by incorporating the information about the local curvature, hence forcing to move further in the direction, in which the gradient is smaller in order to get around the classic "error valley". More so, gradient descent based algorithms like backpropagation despite being popular among researchers are not known to be efficient algorithms due to the fact that the gradient vanishes at the solution. Hessian-based algorithms like the Lavenberg-Marquardt, on the contrary, allow the network to learn more subtle features of a complicated mapping. The training process converges as the solution is approached, because the Hessian does not vanish at the solution.
The  Lavenberg-Marquardt algorithm is basically a Hessian-based algorithm for nonlinear least square optimization [57]. One Step Secant method is an approximation of the Gauss-Newton method for error minimization.  The advantage of this method is the smaller memory requirement and lesser computation time, since unlike other algorithms it does not store the complete Hessian matrix, instead at each training iteration it assumes that the previous Hessian was the identity matrix. This has an added advantage that the new search direction can be found without having to compute the matrix inverse [58]. Higher Order Neuron model [59]  is the one which includes the quadratic and higher order basis functions in addition to the linear basis functions to reduce the learning complexity. 
\par
Neuro-fuzzy systems are models where ANN models are combined with Fuzzy systems  to use the best features of both models. While as ANN's are known to be powerful in reaching a solution, Fuzzy systems have an advantage in comparison to ANN for  explaining the decision rules better [60]. Apart from employing these methods we felt that the study would be incomplete without the use of the comparatively lesser used "backprop - momentum" (backpropagation with momentum term added to the learning rule). The momentum term allows network to respond to local gradient and other trends in the error surface. Without the momentum term network may get stuck in some shallow within the local minima.
\par
It is however important to mention here that for real world problems, the above definitions serve only as a guideline and the actual performance of the ANN models on real world problems does not necessarily follow the above theoretical predictions. Therefore, these varied algorithms under the ANN domain can not be used as off the shelf algorithms until sufficient expertise in the field is obtained.
There are several other issues involved in designing and training a multilayer neural network. These are : (a)	Selecting   appropriate number of  hidden layers in the network; (b) Selecting the number of neurons to be  used in each hidden layer; (c) Finding a globally optimal solution that avoids local minima; 
(d) Converging to an optimal solution in a reasonable period of time; (e) Overtraining of the network and (f)	Validating the neural network to test for overfitting.
\par
While as, a lot of emphasis has been put lately on the use of Random Forest (RF)  technique as an efficient tool for $\gamma$-hadron segregation, we believe that a properly selected and well trained neural net algorithm is equally as efficient for this purpose. The results obtained by [15]  in their study obtained a  Quality factor (QF) of $\sim$ 2.8 and $\sim$ 3.0 for Random Forest   and ANN methods respectively when applied to the MAGIC data. The maximum significance also turns out to be comparable at $\sim$ 8.74$\sigma$ and $\sim$ 8.75$\sigma$ for RF and ANN respectively. In another study  conducted by  Boinee et al. [61] on the MAGIC Cherenkov telescope experiment, detailed comparison of RF, ANN, Support Vector Machines  and Classification Trees have been presented.  While as, the optimized RF technique resulted in a classification accuracy of $\sim$ 81.24 $\%$, the classification accuracy for ANN turned out to be $\sim$81.75$\%$ with a mean error rate of $\sim$0.276 and $\sim$ 0.256 for the Random Forest  and ANN techniques respectively, thereby suggesting that the two techniques are at best comparable. The results obtained from other methods turn out to be quite inferior compared to the ANN and Random Forest, suggesting that both the methods are equally suitable.
\section{Gamma/hadron separation using ANN}

\subsection{ Preparation  of  Training, testing and validation data}
Training the ANN means iteratively minimizing the error between the desired output and the ANN generated value, with respect to the network weights. Clearly, in order for the network to yield appropriate outputs for given inputs, the weights must be set to suitable values. This is done by 'training' the network on a set of input vectors, for which the ideal outputs (targets) are already known. 
For training the ANN we have used $\sim$13750  $\gamma$-ray simulated events following a power law  distribution with a differential spectral index of $\sim$-2.6. This  data-base was obtained   by combining together $\sim$2750 events each at 5 different zenith angles ($\theta$ = 5$^0$, 15$^0$, 25$^0$, 35$^0$ and 45$^0$). 
The cosmic  ray   data   of  $\sim$11290  events,   used  for  training the ANN, is the actual  experimental  data  recorded by the TACTIC telescope and was prepared in the following manner. Around one-third of the data  used ($\sim$ 3163 events) were recorded in the Crab Nebula off source direction. From the Crab Nebula on-source data  base, collected between Nov.10, 2005 - Jan. 30, 2006, we used another ($\sim$ 3163 events)  for which $\alpha\geq 27^\circ$ and are hence certainly cosmic-ray events. The remaining one-third portion of the data  was  taken from $\sim$30h  of  Mrk 421 off-source  observations and this data was collected during the same observing season. The zenith angle of  the off-source observation was restricted to $\leq45^\circ$. The reason for generating the training  data in this manner was to ensure that all  possible systematic influences on the training of the network such as  variable sky brightness in different directions are also included during  the  training procedure. Using the experimental data-base for the protons is a useful way of training, since it helps ANN to recognize the latent patterns, if any, in a better way which can otherwise be difficult to replicate in simulations e.g, in situations  when  the  sky brightness  is higher  than what has been assumed in simulations. The importance of using  real background hadronic events instead of simulated events has also been demonstrated in [14].  
\par
The test data set consists of an independently generated sample of about 44831  events (mixture of $\sim$ 24603 simulated $\gamma$-ray and $\sim$ 20228 actual cosmic-ray  events), which has not been used  while  training  the  ANN.  This data set has exactly the same format as the training data set and is generated in the same manner as the training data.  A validation data sample of   $\sim$ 29798 events ( mixture of 16424 simulated $\gamma$-ray and 13374 actual cosmic-ray  events)  is used for  verifying   that  the network   retains   its    ability  to generalize  and is  not  "over-trained". 
\subsection{ ANN training  and optimizing the number of hidden layer nodes}
The  network  used  in this  work  comprises 6 nodes  in the input layer with one each node for Zenith angle ($\theta$), SIZE, LENGTH, WIDTH, DISTANCE  and  FRAC2 and one neuron in the output layer whose value decides to which class the output is to be  categorized. This value is designated as 0.1 or 0.9 depending upon whether the event in question is a gamma-ray  or a  cosmic-ray  event respectively.  In  order  to determine the optimum number of neurons in the hidden layer  we evaluate the Mean Square Error (MSE) generated by the network. The MSE for the network is defined as:
\par 
\begin{equation}                   
MSE =\frac{1}{2} {\frac{1}{PI}\sum \limits_{p=1}^{P} \sum \limits_{i=1}^{I} \left(\frac{D_{p i} - O_{p i}}{D_{p i}}\right)^2}   
\end{equation}  
\par 
where $D_{pi}$ and $O_{pi}$ are the desired and the observed values and P is the number of training patterns and I is the number of outputs, which happens to be 1 in our case.
Thus MSE defined above, is the sum of the squared differences between  the desired  output and the actual output of the output 
neurons averaged over all the training exemplars [62].  The ANN  algorithms   used  in the   present  work  are  the following: Backpropagation, Resilient Backpropagation, Backprop-momentum, Conjugate Gradient, One step secant, Higher Order Neurons,  Levenberg Marquardt and the Neuro fuzzy.   
\par
With  regard   to  choosing   the  number of  nodes in the  hidden layer, it is well known  that, while  using too few nodes will starve the network  of the resources that it needs  to  solve a particular  problem; choosing   too many nodes has the risk  of potential overfitting  where  the network tends to remember  the training  cases  instead of generalizing  the patterns. In order  to find  the  optimum  number of  nodes in the  hidden layer we employed a  two step procedure.  In the first step we varied  the number of   nodes in the  hidden layer from 5 to 60  (in steps  of  5 up to 40 and in steps of 10 thereafter) and  noted  down  the MSE for  each of the configurations. In the second step,  we  deliberately   used  significantly  higher number of  nodes in   the hidden layer (equal to 90) and then applied  the Singular Value Decomposition (SVD) method  for  identifying  the redundant  nodes [32, 63-65].  It  is  worth mentioning   here  that   determining the optimum number of neurons in the hidden layer by sequentially   increasing  the  number of  nodes  from  60 onwards  involves massive computational effort, hence the need of applying the SVD method is justified. 
\par
In the SVD method, the weight matrix (denoted by F in the present  work) was   generated  by finding   the output of each of the 90 nodes before subjecting  them to the  nonlinear transformation  (i.e  output  of  the hidden node). With a total of 25040  training patterns  and  one  hidden layer  with  90  nodes, the matrix F has thus 25040  rows and 90 columns.  The  SVD  of the matrix  F is given by  F=U S V $^T$, where   U and V  are the  orthogonal  matrices and S is a diagonal  matrix   with  25040 rows and 90 columns.  The  matrix S  contains   the singular  values of  F  on its diagonal.  The dominance  of the  significant singular values of  F  ( say  g out of a  total  p  singular values)  is found  out   by 
using  the  so called  percentage of  energy explained ($P_{ex}$) and is defined  as :
\par
\begin{equation}
P_{ex}=\frac{\sum_{i=1}^{g}{S_i}^2}{\sum_{i=1}^{p}{S_i}^2} \times 100
\end{equation}
where     $S_1, S_2, S_3  ----- S_p $  are the  singular values of  F  arranged  in  their  descending   order [66].  
The  results  of   this study  are  shown in  Fig 3.  where  $P_{ex}$  is plotted   as a function  of  number of nodes in the hidden  layer  for a representative example of 4 ANN  algorithms.  Consolidated  results concerning the performance  of the various  algorithms with regard   to   their   corresponding  MSE  values  for the  training, test  and  validation  data   samples  are  given in Table 5.  
\begin{figure}[h]
\centering
\includegraphics*[width=1.0\textwidth,angle=0,clip]{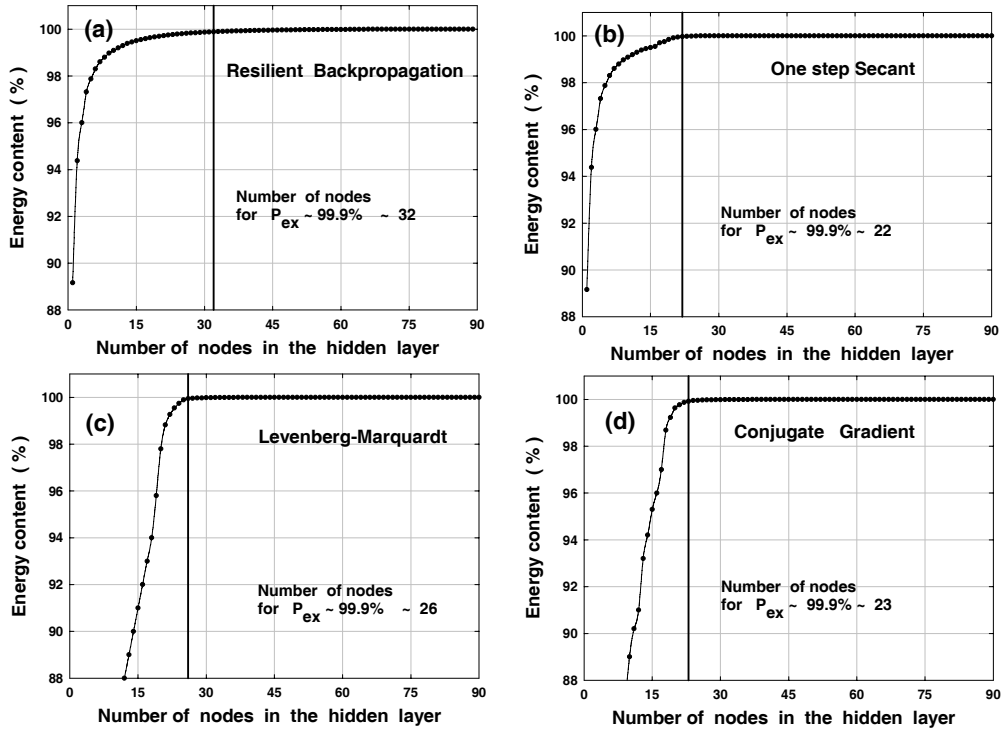}
\caption{\label{fig ---}  Percentage of  energy  explained ($P_{ex}$) as a function of  number of nodes in the  hidden layer   for some representative algorithms : (a) Resilient backpropation  (b)  One Step Secant  (c)  Levenberg-Marquardt algorithm and  (d) Conjugate Gradient. The optimum  number of nodes for $P_{ex}$ $\sim$ 99.9 $\%$ is also marked in the figures by full vertical lines.} 
\end{figure} 
The results presented in this  table shown separately for 35 and 90 nodes in the hidden layer, can be used for checking whether the ANN algorithm is "over-trained" or not. When the network is over-trained, the MSE for the test and validation data samples are expected to be significantly higher than the corresponding value of MSE achieved during training.   
\\
\begin{table}[h]
\begin{center}
\caption{MSE  values  of  various ANN algorithms   for  the  training, test and  validation data  samples.  The  two values  presented  in the table correspond to 35 and 90 nodes in the hidden layer.}
\begin{tabular}{|c|c|c|c|}
\hline
$	Algorithm              $  &   $Train \ 35/90   $ &  $ Test \ 35/90   $  & $ Valid \ 35/90     $ \\
\hline 
$Backpropagation         $  &   $0.103/0.102   $ &  $ 0.103/0.103  $  & $ 0.103/0.103     $ \\ 
\hline 
$Backprop Momentum       $  &   $0.156/0.158   $ &  $ 0.157/0.159  $  & $ 0.156/0.158     $ \\
\hline 
$Resilient Backprop      $  &   $0.035/0.033   $ &  $ 0.036/0.035  $  & $0.036/0.034     $ \\
\hline 
$ Scale Conjugate        $  &   $0.047/0.040   $ &  $ 0.046/0.041  $  & $0.047/0.041     $ \\
\hline
$ One Step Secant				 $  &   $0.053/0.050   $ &  $ 0.053/0.051  $  & $0.053/0.051      $ \\
\hline
$Lavenberg Marquardt     $  &   $0.017/0.015   $ &  $0.017/0.030   $  & $0.017/0.031      $ \\
\hline 
$Higher Order           $  &   $0.039/0.033   $ &  $  0.040/0.033 $  & $0.040/0.034      $ \\
\hline
$Neuro Fuzzy            $  &   $0.062/0.062   $ &  $  0.062/0.063 $  & $0.062/0.062     $ \\
\hline
\end{tabular}
\end{center}
\end{table}
\\
The  optimum  number of nodes   for   $P_{ex}$ $\sim$ 99.9 $\%$  is  also marked in the  figures  by   full  vertical lines. 
For   $P_{ex}$ $\sim$ 99.9 $\%$,  one  can  easily find  from  the   this  figure  that the optimum  number of nodes needed 
for obtaining   the desired  results  varies  between $\sim$22 to  $\sim$ 32.  Except for  the Backpropagation-Momentum  algorithm which requires  only  $\sim$5  nodes,   the  remaining  algorithms   are  also  found   to   yield  optimum  performance with $\sim$20 to $\sim$30  nodes  in the hidden layer.  The reason for Backprop momentum requiring too few nodes can be understood from the manner in which the algorithm is trained. In this algorithm, momentum term is added to the Backprop to enhance the training time with a slight compromise on the performance of the network. This effect is seen in our case also where we see the Backprop Momentum algorithm yielding the worst MSE value compared to all other algorithms.
\par
On  examining   Fig.3 and  Table 5 one  can   arrive at the following  conclusions :
(i) None  of the ANN algorithms used in this  work  are  under trained or over trained  if about 35 nodes are used in the hidden layer.  (ii) Increasing  the  number  of nodes  beyond  35 results in only a marginal reduction in the MSE. (iii) The MSE value yielded by the Levenberg-Marquardt method  with 35 nodes is found out to be the lowest compared  to all other  ANN algorithms.  (iv) Increasing the number of nodes  from 35 to 90 leads to the problem of overfitting in the Levenberg-Marquardt method. (v) For the remaining algorithms no overfitting  problem  is seen when 90 nodes  are used in the hidden layer. The overfitting of  the Levenberg-Marquardt (with 90 nodes in the  hidden layer) is  most probably  related to the way in which the training is performed in this algorithm, more specifically  how the algorithm accounts for error as well as  the gradient information  based on  blending  between the gradient descent  method  and the Gauss Newton rule.  The Levenberg-Marquardt trains  in such a way that large  steps are  taken in the direction  of  low curvature  to skip past  the plateaus quickly, and  smaller steps are taken in the  direction  of  high curvature  to slowly  converge  to the global  minima. Thus every narrow valley or plateau, even if as a result of noise in the data, is important for this method.  Hence, when larger  number  of nodes are  presented ( i.e, 637 weights for the 90 nodes versus  252 weights  for the 35  nodes  in the hidden layer), the algorithm  becomes  sensitive  even  to  the noise  values  present  in the  data,  which with  lesser  number of nodes  could  have been ignored. The source of noise  in our  training/test  data-base is as  result of  inherent  fluctuations  in  the  shower  development  process.  On the basis of the above argument one can thus safely use 35 nodes in the hidden layer for all the algorithms. 
\par
It is worth mentioning here that the  modification  of the ANN  structure  by  analyzing  how  much each  node  contributes to the actual output  of the neural network and  dropping  the  nodes  which do not  significantly affect  the output  is  also referred  to as  pruning. The  basic  principle  of  pruning  relies  on the fact  that if  two hidden nodes give the same outputs for  every input vector, then   the  performance of the neural network  will not be   affected  by removing  one of  the nodes in the hidden layer. In  the  SVD approach, redundant  hidden  nodes cause singularities in the weight matrix which can be identified  through  inspection of its  singular  values.  A non-zero  number of small singular values indicates redundancy in the initial choice  for the number of  hidden layer  nodes  and  the  approach  can be  safely  used  for  eliminating  these  nodes  to attain  the pruned  network model.  
\par
A plot of the mean square error as a function of the number of nodes in the hidden layer for the most popular standard  backpropagation  network     and the Lavenberg-Marquardt algorithm with  Sigmoid  transfer function is shown in  Fig.4a.  
\begin{figure}[h]
\centering
\includegraphics*[width=1.0\textwidth,angle=0,clip]{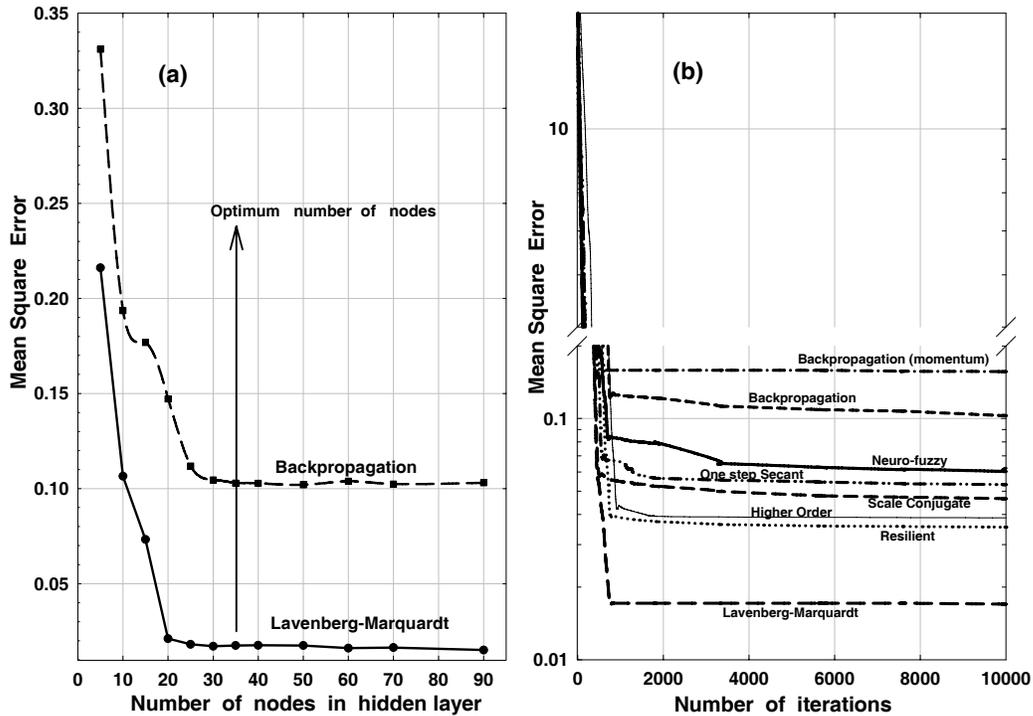}
\caption{\label{fig ---} (a)  Mean  square error  as a function of   number of nodes in the  hidden layer for the  Backpropagation  and the Lavenberg-Marquardt algorithms. (b)  Mean square  error  for   various  ANN algorithms  as a function of number  of  iterations  with  35 nodes in the hidden layer.} 
\end{figure} 
While the MSE at the end of the training, for  35 nodes  in the hidden  layer,  is $\sim$0.1032 for the backpropagation network, the corresponding vaule for the  Lavenberg-Marguardt algorithm is found to be $\sim$0.0171. Although the MSE yielded by the Lavenberg-Marguardt algorithm is found to be lower than the MSE values of other training algorithms, including the backpropagation algorithm, the reason for showing the MSE for the backpropagation algorithm is mainly because it has been considered as a "work-horse" in the field of neural computation. 
\par
The  variation of the MSE as a function of number of iterations  for  all ANN algorithms used, is shown in Fig.4b. The number of neurons in the  hidden layer was thus fixed  at 35 nodes for all these algorithms. In all above algorithms, the training is continued till the MSE error  reaches a plateau and does not decrease any further. About 10,000 iterations were generally found to be sufficient to train the ANN on various algorithms.  This superior convergence of Lavenberg-Marquardt algorithm over the conventionally used backpropagation algorithm and/or resilient backprop is not totally unexpected and has been demonstrated by us on standard benchmark and regression problems [60].   
\par 
It is worth mentioning here that for studying the performance of the various ANN algorithms we have used BIKAS (BARC-IIT Kanpur ANN Simulator) ANN package [67] and MATLAB [68,69]  neural net packages. While as, MATLAB has been used for backpropagation, resilient backpropagation, Scale Conjugate, backprop-momentum, Lavenberg-Marquardt, and One Step Secant algorithms, the BIKAS package has been used for Higher Order Network and Neuro-Fuzzy models. 
\subsection{Testing and validation of Lavenberg-Marquardt ANN algorithm}
Since MSE error returned by the Lavenberg-Marquardt algorithm is lower  than  the  MSE error values of  other methods including the backpropagation method, we have used  only  this algorithm  on the test data set for  a  more descriptive  analysis.   When the test data-base is presented to the network, instead of yielding the desired output as 0.1 or  0.9,  the ANN outputs a range of values between 0.1 to 0.9.  The broad distributions around 0.1 and 0.9, returned after testing the prior trained ANN algorithm, instead of the desired 0.1 or 0.9, is on account of the inherent shower to shower fluctuations on an event to event basis even though train and test data is generated in a similar manner.  The response of the network (i.e., frequency distribution of the selected events) for the test data sample comprising simulated $\gamma$-rays  and actual  background  as  a function of the  ANN output is shown  in  Fig.5a.   
\begin{figure}[h]
\centering
\includegraphics*[width=1.0\textwidth,angle=0,clip]{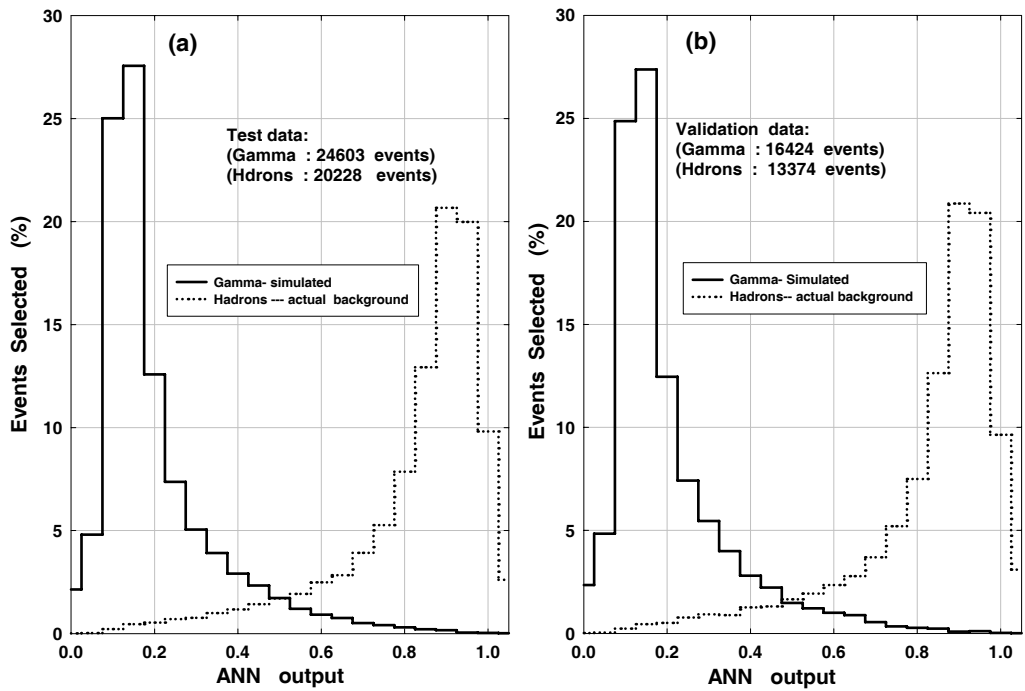}
\caption{\label{fig ---} (a)  ANN output of  Lavenberg-Marquardt algorithm  in response to  simulated $\gamma$-rays  and  actual  background events  of  the  test data sample comprising   a  total  44831  events.  
(b) Same as (a) except for an independant validation data sample comprising a total 29798 events. No cut on $\alpha$ has been applied  to the data  presented in these  figures.} 
\end{figure} 
The results  obtained  for the validation data sample are  shown in Fig. 5b.  Excellent  matching of the  results  obtained  for  the  test  and  validation data clearly demonstrates  that ANN  has indeed "learned" and simply not remembered the classification. It is important  to mention here  that   no cut on $\alpha$ has been applied  to the data  presented in these  figures. 

\section{Determination  of   optimum  ANN cut value}

For   determining  the   ANN output cutoff value ($\eta_{cut}$),  which  will optimize the  separation of the two  event  classes (i.e $\gamma$-ray and cosmic-rays),  one  can   maximize  either  Quality Factor ($QF$)  or  more adequately,    statistical  significance 
($ N_{\sigma}$). Following  their  standard  definitions [15],  these are  given by :
\par
\begin{equation}
QF=\frac {N_{\gamma}/N_{\gamma0}} {\sqrt { N_{p}/N_{p0}}}=  \frac {f_{\gamma}}  { \sqrt{ f_p}} \\
\end{equation}
\begin{equation}
N_{\sigma}=\frac {N_{\gamma}} {\sqrt {  N_{\gamma}+ 2 N_p}}
\end{equation}
\par
where  $N_{\gamma}$  and   $N_{p}$  are the  number of $\gamma$-rays and  hadrons, respectively, after classification;
 $N_{\gamma0}$  and   $N_{p0}$  are the  number of $\gamma$-rays and  hadrons, respectively, before  classifier  and    
$f_{\gamma}$ and   $ f_p$   are the corresponding  acceptances for  $\gamma$-rays  and hadrons.   
Although  many   groups   have  used  QF  for   optimizing   the  performance  of  their  classification methods [14] , we have 
optimized  the performance of the ANN  on the basis of  maximizing $N_{\sigma}$.   The reason  for this  is  the  fact  that a high value of $QF$ can also  result  from   tight  cut which  can  reduce  the   $\gamma$-ray retention  capability  of the   classification method.   Furthermore,  maximization  of    $N_{\sigma}$    also  ensures   that classification  procedure  in not    biased  unfavorably   towards   higher  energies.  Optimization  on the  basis of  maximizing $N_{\sigma}$  has also been followed  by  other  groups [13,15].
\par
It is worth mentioning   here  that  definition  of  statistical  significance 
($ N_{\sigma}$) given above   can be  only used  when  $N_{\gamma}$  is  known beforehand  which is  possible  only when one is dealing    with   simulated data.  Since, in   case of  actual data  collected with  Cherenkov imaging telescopes,  $N_{\gamma}$  can also be calculated statistically by subtracting the expected number of background events ( e.g  $27^\circ\leq\alpha\leq81^\circ$  used  by us in [39] and in this work) from the  $\gamma$-ray domain events (e.g $\alpha\leq18^\circ$  in our case)   the  definition  of  statistical  significance given  above   needs  to  be  modified. While we have used the  above  expression  of   $ N_{\sigma}$    for estimating  $\eta_{cut}$,   the significance  of the  $\gamma$-ray events  found   in the  actual  Crab Nebula data  has been calculated  by following   a more  rigorous method  of  using  maximum likelihood ratio of Li and Ma [70].
\par 
The  value  of $\eta_{cut}$  defines the decision boundary between the two event species  and  in order   to determine   its   optimum value  we  used  a  data sample  of about 12953 events (mixture of 8865 simulated $\gamma$-ray and 4088 actual cosmic-ray  events). The zenith angle range of these events was again chosen to be in the range (0-45)$^\circ$. Since  the  value  of   $N_{\sigma}$   depends critically on the number of  $\gamma$-rays present in the data  we have considered  only  $N_{\gamma0}$$\sim$177 ( i.e $\sim$2$\%$ of the total  $\gamma$-rays  present  in the data sample) for  determining  the  optimum value  of  $\eta_{cut}$. The  event  is  classified  as a $\gamma$-ray  like event  only if the corresponding  ANN output ($\eta$)  is $\leq$ $\eta_{cut}$   and  $\alpha\leq 18^\circ$.   The calculation  was  performed   by varying  $\eta_{cut}$  from 0.05 to 1.0 in steps of 0.05 and   recording    $N_{\sigma}$  at each value of $\eta_{cut}$. The  results  of this  study  are given in Fig.6  which shows  variation  of   $N_{\sigma}$  as  a function of  $\eta_{cut}$  for the Levenberg-Marquardt based  ANN algorithm. 
\begin{figure}[h]\centering
\includegraphics*[width=1.0\textwidth,angle=0,clip]{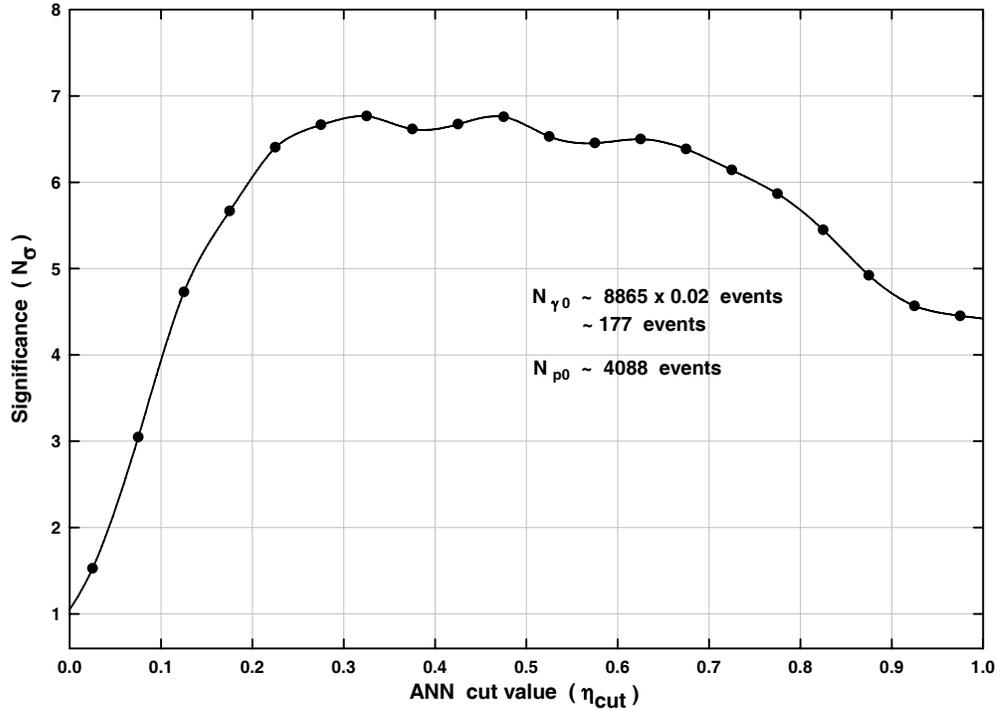}
\caption{\label{fig ---}  Variation   of   statistical  significance 
($ N_{\sigma}$) as  a function  of  ANN cut value ( $\eta_{cut}$) for  the  Levenberg-Marquardt algorithm.} 
\end{figure}
On  examining   this figure  one can  see that maximum  value  of   $N_{\sigma}$ $\sim$ 6.8$\sigma$  is obtained at $\eta_{cut}$ $\sim$0.475.
The above data  has also  been  used for evaluating  the  performance  of  other  ANN  algorithms  and  finding their optimum $\eta_{cut}$ values. The results of this  study  are  summarized in Table 6  where,   in addition to  $N_{\sigma}$  values  yielded  by different  algorithms, we also give the corresponding  $\eta_{cut}$  range  within which   $N_{\sigma}$   stays constant. 
 The   lower  value  of  $\eta_{cut}$  defines  the   tight cut   and  higher  value  designates  the  loose cut.
\\ 
\begin{table}[h]
\begin{center}
 \caption{ Maximum value  of   the  statistical significance  $N_{\sigma}$   yielded  by various  ANN algorithms along with corresponding    $\eta_{cut}$  range with in which   $N_{\sigma}$   stays constant. The   lower  value  of  $\eta_{cut}$  defines  the   tight cut   
 and  higher  value  defines the  loose cut.}
\begin{tabular}{|c|c|c|}
\hline
$	Algorithm		             $    &   $ \eta_{cut} $   &   $   N_{\sigma}    $  \\ 
\hline
$ Backpropagation          $    &   $   0.40 - 0.67   $   &   $   5.21     $  \\
\hline 
$Backpropagation \ monemtum$    &   $   0.30 - 0.57   $   &   $   5.33     $ \\
\hline
$ Resilient\ Backprop      $    &   $   0.42 - 0.67   $   &   $   5.25     $ 	\\
\hline
$ Scale \ Conjugate        $    &   $   0.40 - 0.67   $   &   $   4.80     $ 	\\
\hline
$ One \ Step \ Secant      $    &   $   0.42 - 0.67   $   &   $   5.25     $ 	\\
\hline
$ Lavenberg \ Marquardt    $    &   $   0.30 - 0.62   $   &   $   6.80     $ 	\\
\hline
$ Higher \ Order           $    &   $   0.40 - 0.70   $   &   $   4.80     $ 	\\
\hline
$ Neuro \ Fuzzy            $    &   $   0.40 - 0.67   $   &   $   4.47     $ 	\\
\hline
\hline
$ Dynamic \ Super \ cut    $    &   $       ---       $   &   $   6.09     $ 	\\
\hline
\end{tabular}
\end{center}
\end{table}
\\
The  value of   $N_{\sigma}$   achieved  with  Dynamic Supercuts  is  also  shown in the table  for comparison.
It is quite evident from the table    that  out of 8 different  ANN algorithms studied here,  Levenberg-Marquardt algorithm     yields  the best results.  The value  of  $N_{\sigma}$   for  other  algorithms   is  found to vary  from  $\sim$ 4.5$\sigma$  ( Higher order network)  to   $\sim$ 5.3$\sigma$  (backprop-momentum). Because  of the  superior  performance  of  the    Levenberg-Marquardt algorithm, we will  only use this algorithm  for analyzing  the  actual Crab Nebula data.  
\par
Referring  back to Fig.6,  since   the  change  in   $N_{\sigma}$  is   insignificant  when   $\eta_{cut}$ is varied  from 
$\sim$0.3 to  $\sim$0.5,  we will use  a value  of  $\eta_{cut}$ $\sim$0.5  for   analyzing  the  actual Crab Nebula data. Admittedly,  using  $\eta_{cut}$ $\sim$0.5   also  increases the cosmic  ray background.
The reason for choosing   the  higher  $\eta_{cut}$ value   is  to ensure   that  we retain   maximum  number of $\gamma$-rays  from  the  source.  For  sources  which  are  weaker  than  the  Crab Nebula   one  can use  $\eta_{cut}$ $\sim$0.3 
 so that  contamination  from  more  background  can be  reduced.    Since   our  main   preference   is  to   observe  relatively   stronger sources such as blazars  using   $\eta_{cut}$ $\sim$0.5  is   an obvious  choice  if  we  want     to   measure  their   energy  spectra   beyond  energies  of $\sim$10TeV.  Following this  approach of choosing the tight cuts for detecting weaker/new sources and loose cuts for obtaining the energy spectrum, is a well known procedure which is adopted by almost all the groups who  work 
on  Cherenkov imaging  telescopes.     


\section{Application  of   the  ANN  methodology to the  Crab Nebula data collected  with the TACTIC  telescope}

In order to study the  $\gamma$/hadron  segregation  potential  of the ANN  methodology, we  have  applied  this selection method to  the  Crab  Nebula  data  collected  with   the TACTIC telescope.  
For this purpose  we  reanalyzed   the   Crab Nebula  data for $\sim$101.44 h collected during  Nov. 10, 2005 - Jan. 30, 2006. The zenith angle during the observations was $\leq$45$^\circ$ and  the  data was collected with inner 225 pixels ($\sim$ 4.5$^\circ$ $\times$ 4.5$^\circ$)  of the full imaging camera  with the innermost 121 pixels 
($\sim$ 3.4$^\circ$ $\times$ 3.4$^\circ$) participating  in the trigger.  The standard two-level image  'cleaning' procedure  with picture and boundary thresholds of  6.5$\sigma$ and 3.0$\sigma$, respectively  was employed to obtain the clean Cherenkov  images. Details of this analysis procedure and the data collecting methodology for this period can be found in [42].  The purpose of this image cleaning procedure is to take care of the fluctuations in the image which arise due to electronic noise and night sky background variations. These clean Cherenkov images were then characterized by calculating their standard image parameters like LENGTH, WIDTH, DISTANCE, $\alpha$, SIZE and   FRAC2.  Before  investigating   the  $\gamma$/hadron  segregation  potential  using ANN  methodology, we will 
first  apply   the  standard   Dynamic  Supercuts  procedure  [8]  to  the   data  for  extracting   the   $\gamma$-ray signal  from  the  background  cosmic-ray events. 
\par
The  cut  values used  for the analysis  are the  following :    $0.11^\circ\leq  LENGTH \leq(0.260+0.0265 \times \ln S)^\circ $,   $0.06^\circ \leq WIDTH \leq (0.110+0.0120 \times \ln S)^\circ$,
$0.52^\circ\leq DISTANCE \leq 1.27^\circ cos^{0.88}\theta$, $SIZE \geq 450 d.c$ ( where 6.5 digital counts$\equiv$1.0 pe ),
$\alpha \leq 18^\circ$  and $FRAC2 \geq 0.35$. It is important  to emphasize  here  that  the  Dynamic Supercuts  $\gamma$-ray selection  criteria  used in the present analysis are  the same  which  we had  used  in our previous  work [39] for  developing  an ANN-based energy reconstruction procedure  for the TACTIC  telescope. Since  the present work  uses the same  data-base  as well as the same  energy reconstruction  procedure, we will consider the previous work [39] as some sort of benchmark for the present study. 
Admittedly,    there   may be   a  scope   for   optimizing  the   previously  used  Dynamic Supercuts   further   (e.g  by  using  cuts  which depend on both energy   and  zenith angle),  but  the  results  of this  study will be presented   elsewhere.   
\par
A well established  procedure to extract the $\gamma$-ray  signal from the cosmic-ray background using a single imaging telescope is to plot the  frequency distribution of  $\alpha$  parameter which is expected to be flat  for the isotropic background of cosmic events [8]. For $\gamma$-rays, coming from a point source, the distribution is expected to show a peak  at  smaller $\alpha$ values. Defining $\alpha\leq18^\circ$ as the $\gamma$-ray domain and $27^\circ\leq\alpha\leq81^\circ$ as the background region, the number of $\gamma$-ray events is then calculated  by subtracting the expected number of background events (calculated  on the basis of background  region) from the  $\gamma$-ray domain events. 
The number of $\gamma$-ray events  obtained after applying  the above cuts are found to be $\sim$(928$\pm$100) with a statistical significance of $\sim$9.40$\sigma$.  The significance  of the excess  events  has  been   calculated by using the maximum  likelihood ratio method of Li $\&$ Ma [70]. The $\alpha$-distribution is given in Fig. 7a and the corresponding  differential energy spectrum  of the Crab Nebula   shown  in Fig. 7b,  has  been  computed using the following formula: 
\begin{figure}[h]
\centering
\includegraphics*[width=1.0\textwidth,angle=0,clip]{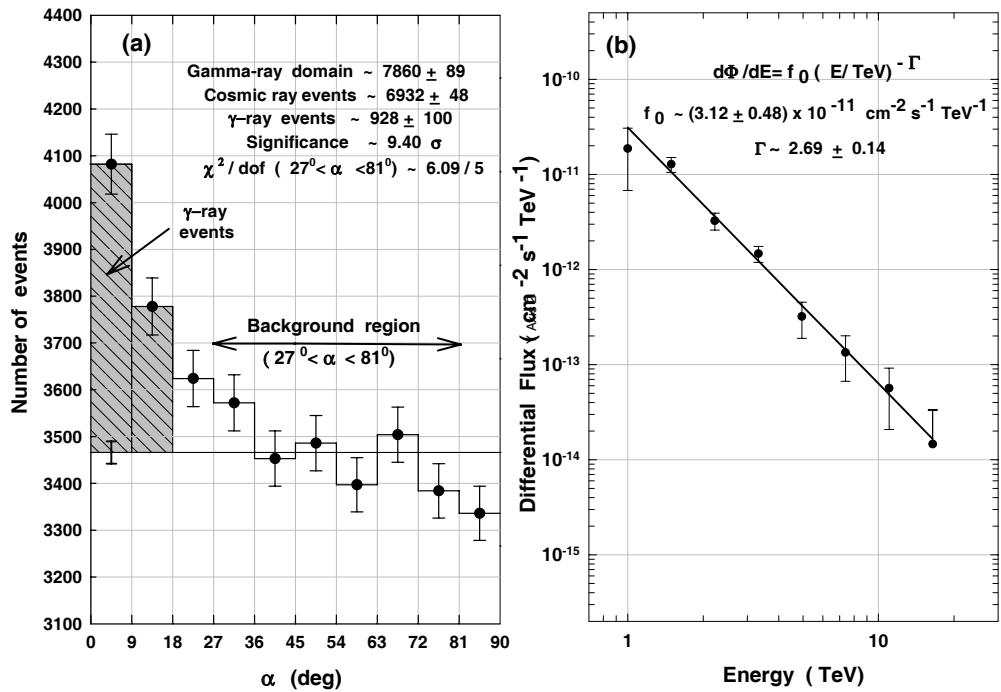}
\caption{\label{fig ---} (a)  Crab Nebula  $\alpha$-plot  for $\sim$101.44 h  of  data  using  Dynamic Supercuts $\gamma$-ray selection  criteria. (b)  The corresponding  differential  energy spectrum  of the Crab Nebula as measured by the TACTIC  telescope.} 
\end{figure} 
\par
\begin{equation}
\frac{d\Phi}{dE}(E_i)=\frac {\Delta N_i}{\Delta E_i \sum \limits_{j=1}^5 A_{i,j} \eta_{i,j} T_j}
\end{equation}
\par
where $\Delta N_i$ and $d\Phi(E_i)/dE$ are the number of events and the differential flux at energy $E_i$, measured in the ith  energy bin $\Delta E_i$ and over the zenith angle range of 0$^\circ$-45$^\circ$, respectively. $T_j$ is the observation time in the jth zenith angle bin with corresponding energy-dependent effective area ($A_{i,j}$) and $\gamma$-ray acceptance ($\eta_{i,j}$). The 5 zenith angle bins (j=1-5) used are 0$^\circ$-10$^\circ$, 10$^\circ$-20$^\circ$, 20$^\circ$-30$^\circ$, 30$^\circ$-40$^\circ$  and 40$^\circ$-50$^\circ$ with  effective  collection area  and  $\gamma$-ray acceptance  values   available at 5$^\circ$, 15$^\circ$, 25$^\circ$, 35$^\circ$ and 45$^\circ$. The number of $\gamma$-ray events  ($\Delta N_i$)  in a particular  energy bin is  calculated  by subtracting the expected number of background events, from the  $\gamma$-ray domain events. 
The $\gamma$-ray differential  spectrum, shown  in Fig. 7b,   has  been   obtained   after  using  appropriate values of  effective collection area and $\gamma$-ray acceptance  efficiency  (along with  their  energy and zenith angle dependence).  A power law fit  $(d\Phi/dE=f_0 E^{-\Gamma})$  with  $f_0 \sim (3.12\pm0.48)\times 10^{-11} cm^{-2}s^{-1}TeV^{-1}$  and 
$\Gamma \sim2.69\pm0.14$ is  also shown in Fig 7b. The fit  has a $\chi^2/dof\sim3.64/6$ with a corresponding probability of $\sim$0.72. Details of the energy reconstruction procedure can be seen in [39] which uses 3:30:1 ANN configuration with SIZE, DISTANCE and Zenith angle as the inputs to the neural net.
\par
While applying the already trained Lavenberg-Marquardt based ANN network, with 6:35:1 configuration,  for extracting the $\gamma$-ray signal from the data, the number of $\gamma$-ray events are found out to be $\sim$(1141$\pm$106) with a statistical significance of $\sim$11.07$\sigma$.
A  value  of  $\eta_{cut}$ $\sim$0.50 has  been  used  for  selecting  $\gamma$-ray  events  and  only those events are allowed to go for classification with ANN, which satisfy the prefiltering cuts (SIZE $\geq$ 50pe and $0.4^\circ$ $\leq$ DISTANCE $\leq$ $1.35^\circ$). The $\alpha$-distribution of  the   ANN  selected  events is given in Fig.8a, while as the corresponding  differential  energy spectrum  is shown  in Fig.8b.  
\begin{figure}[h]
\centering
\includegraphics*[width=1.0\textwidth,angle=0,clip]{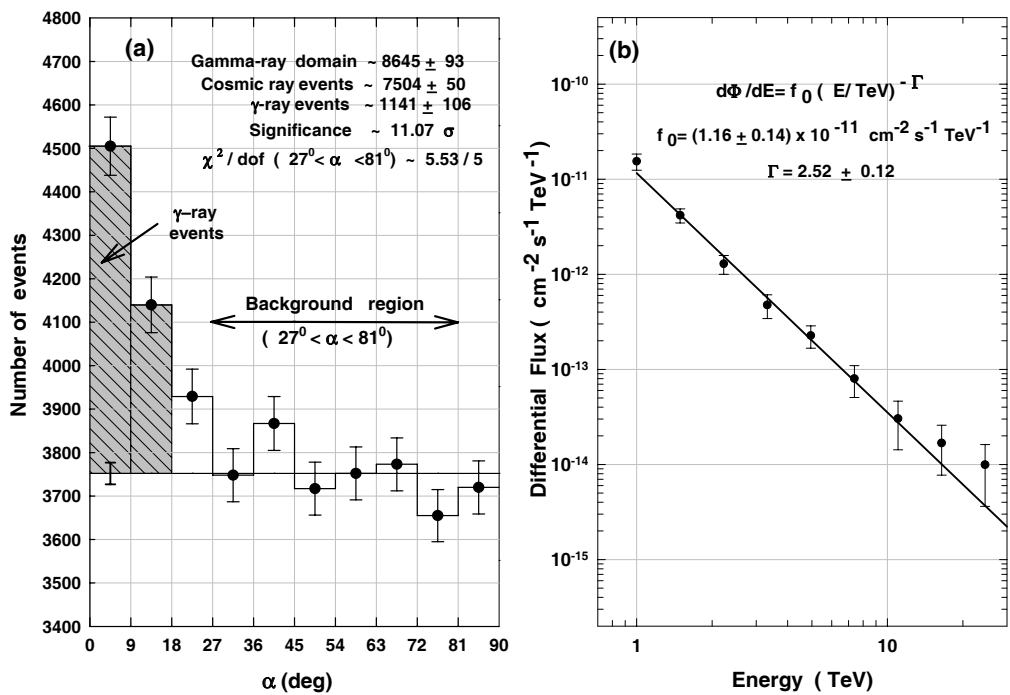}
\caption{\label{fig ---} (a)  Crab Nebula  $\alpha$-plot  for $\sim$101.44 h  of  data  using Lavenberg-Marquardt based ANN network $\gamma$-ray selection  criteria. (b)  The corresponding  differential energy spectrum  of the Crab Nebula  when  ANN network is  used  for   selecting  $\gamma$-ray  like   events.} 
\end{figure} 
A power law fit  $(d\Phi/dE=f_0 E^{-\Gamma})$  with  $f_0 \sim (1.16\pm0.14)\times 10^{-11} cm^{-2}s^{-1}TeV^{-1}$  and $\gamma \sim2.52\pm0.12$ is also shown in Fig 8b. The fit  has a $\chi^2/dof\sim 4.58/7$ with a corresponding probability of $\sim$0.71. Reasonably good matching of the Crab Nebula spectrum with that obtained by the Whipple and HEGRA groups [71,72] reassures that the procedure followed by us for selecting  $\gamma$-ray like events as well as obtaining the energy spectrum of a source, is quite reliable. 
\par  
On comparing  the results  of  Dynamic Supercuts $\gamma$-ray selection  procedure (Fig.7) with the  Lavenberg-Marquardt based ANN network (Fig.8) it is evident that the performance of the later is somewhat superior,  both with  regard  to improving  the  statistical significance of the $\gamma$-ray signal as well as in selecting more number of $\gamma$-rays. Although the   improvement  (i.e  gain of $\sim$213 gamma-ray like events along with signal enhancement from 9.4$\sigma$ to 11.07$\sigma$) looks to be only modest, the main advantage accruing from the ANN methodology is that it is more efficient at higher energies which has allowed us to extend the Crab Nebula energy spectrum up to an energy  of $\sim$24TeV. At $\gamma$-ray  energies above $\sim$9 TeV, the Lavenberg-Marquardt based ANN network selects  $\sim$(85$\pm$28) events as  against $\sim$(24$\pm$9) events selected by the Dynamic Supercuts procedure. 
\par
When  a value  of  $\eta_{cut}$ $\sim$0.30 is  used,   the number of $\gamma$-ray events are  found out to be $\sim$(680$\pm$67) with a statistical significance of $\sim$10.49$\sigma$ and this is in perfect agreement with the discussion presented in Section 8. Although the use of tight cut   (i.e $\eta_{cut}$ $\sim$0.3) yields almost same statistical significance (ignoring slight degradation) as compared to $\eta_{cut}$ $\sim$0.5 cut case, the number of $\gamma$-rays  retained are significantly less and it is just for this reason that we preferred to use a  somewhat loose cut $\eta_{cut}$ $\sim$0.5.  
\par
The performance  of the Lavenberg-Marquardt based ANN network  was  further  validated  by  applying  it  $\sim$ 201.72  hours of on-source data  collected on  Mrk 421  with  the  TACTIC  telescope  during  Dec. 07, 2005  to Apr. 30, 2006.    The  total data   used here  also includes  observations  from    Dec. 27, 2005 to Feb. 07, 2006   when   the   source  was  found  to be  in a  high state  by the TACTIC  telescope as compared to the rest of the observation period [42]. When already trained ANN is  used for extracting the $\gamma$-ray signal from the data, the number of $\gamma$-ray events are found out to be $\sim$(1493$\pm$121) with a statistical significance of $\sim$12.60$\sigma$.  On comparing  these  results  with that obtained  by  using  Dynamic Supercuts [42] which  yields,  $\sim$(1236$\pm$110) $\gamma$-ray events with a statistical significance of $\sim$11.49$\sigma$, it is  reassuring to find that the ANN method is indeed more efficient than the Dynamic Supercuts method. Furthermore, as expected, no signature of a $\gamma$-ray signal is seen when the ANN method is applied to $\sim$ 29.65 hours of  off-source data. The  results  obtained  with  the  ANN method ( $\sim$ 60$\pm$42 with a statistical significance  of $\sim$1.46$\sigma$) compare well with  the results reported by us earlier using  Dynamic Supercuts [42]( $\sim$ 28$\pm$20 with a statistical significance  of $\sim$0.71$\sigma$).
Detailed  results of the reanalysis using the ANN including the  energy spectrum Mrk-421 will be presented elsewhere.
\par
Successful detection of $\gamma$-rays from Mrk-421 thus clearly demonstrates  the capability of the properly trained ANN to extract a $\gamma$-ray  from a  source  other than the Crab Nebula. It also  indicates   that the generalization  capability  of the   ANN  can be   enhanced  if  it is   trained  with  the  experimental data  collected  from   different  directions  having somewhat  variable sky brightness.          

\section{Comparison of Dynamic Supercuts  and  ANN  analysis  methods}

A detailed  study  for comparing   the performance of  Dynamic Supercuts  and  ANN  analysis  methods  has  also been  conducted  by us so that  the  overall   $\gamma$-retention capability  of  the Dynamic Supercuts  and  ANN  analysis  methods  can be compared.  
One of   the ways to  study  this is  to   use  the  Monte Carlo simulated  data for $\gamma$-rays  and plot  the   dependence of  effective collection areas  as a function of  primary  energy  for  the two  $\gamma$-ray selection methodologies. The results  of this study  are  shown in Fig.9 where effective collection areas   for  the two  $\gamma$-ray selection methodologies is  plotted as a function of energy for two representative   zenith angle  values    of   15$^\circ$ and   35$^\circ$.   
\begin{figure}[h]
\centering
\includegraphics*[width=1.0\textwidth,angle=0,clip]{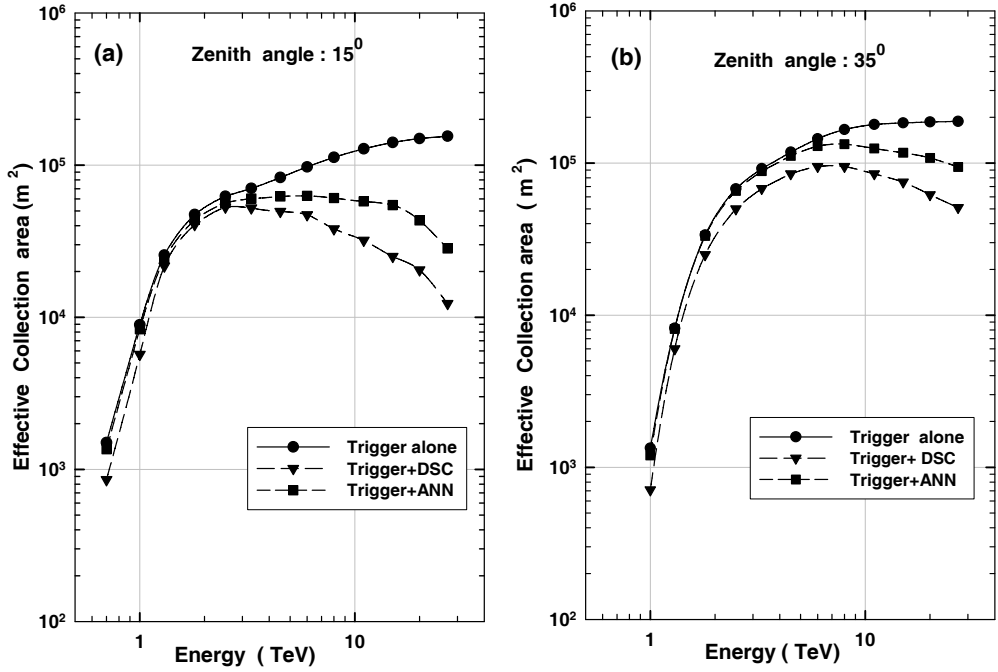}
\caption{\label{fig ---}  Effective collection area as a function  of the primary  $\gamma$-ray energy  for simulated $\gamma$-ray  at showers   zenith  angles of   (a)   15$^\circ$  and (b)35$^\circ$.   While   top most curve (labeled as  Trigger alone)  gives  the effective area    when  no cuts are applied to  the  data,  the   remaining  2 curves  (labeled as  Trigger+ DSC  and  Trigger+ANN)  represent  when Dynamic Supercuts  and  ANN  analysis  methods, respectively  are  applied to the  data}.  
\end{figure} 
Apart  from  showing   the  effective  areas ( i.e $A_{\gamma}(E)f_{\gamma}(E)$)  for the two  $\gamma$-ray selection methodologies, the corresponding  effective area when no cuts  are applied  to  the  data ( i.e $A_{\gamma}(E)$)  is also shown for comparison. 
The results  displayed  in the figure  clearly indicate  that  the efficiency  of  Dynamic Supercuts   is  biased  towards  lower  energies ( particularly at lower  zenith angles).  On the other hand,  it  is  the superior  performance  of  Lavenberg-Marquardt based ANN network  ( i.e  more   collection area  at higher energies) which  has  enabled us  to  retain  relatively higher number of events at energies above $\sim$9 TeV in the actual data as compared to the Dynamic Supercuts  procedure. 
\par
The above conclusion has been further validated by obtaining scatter plots of various  image parameters and  the   results  of this study  are  shown in Fig.10.  This  figure displays scatter plots of LENGTH, WIDTH, DISTANCE and FRAC2 as a function SIZE  for $\sim$8358 events which have been characterized as $\gamma$-ray like  by the  ANN algorithm and   have  $\alpha\leq18^\circ$.  
\begin{figure}[h]
\centering
\includegraphics*[width=1.0\textwidth,angle=0,clip]{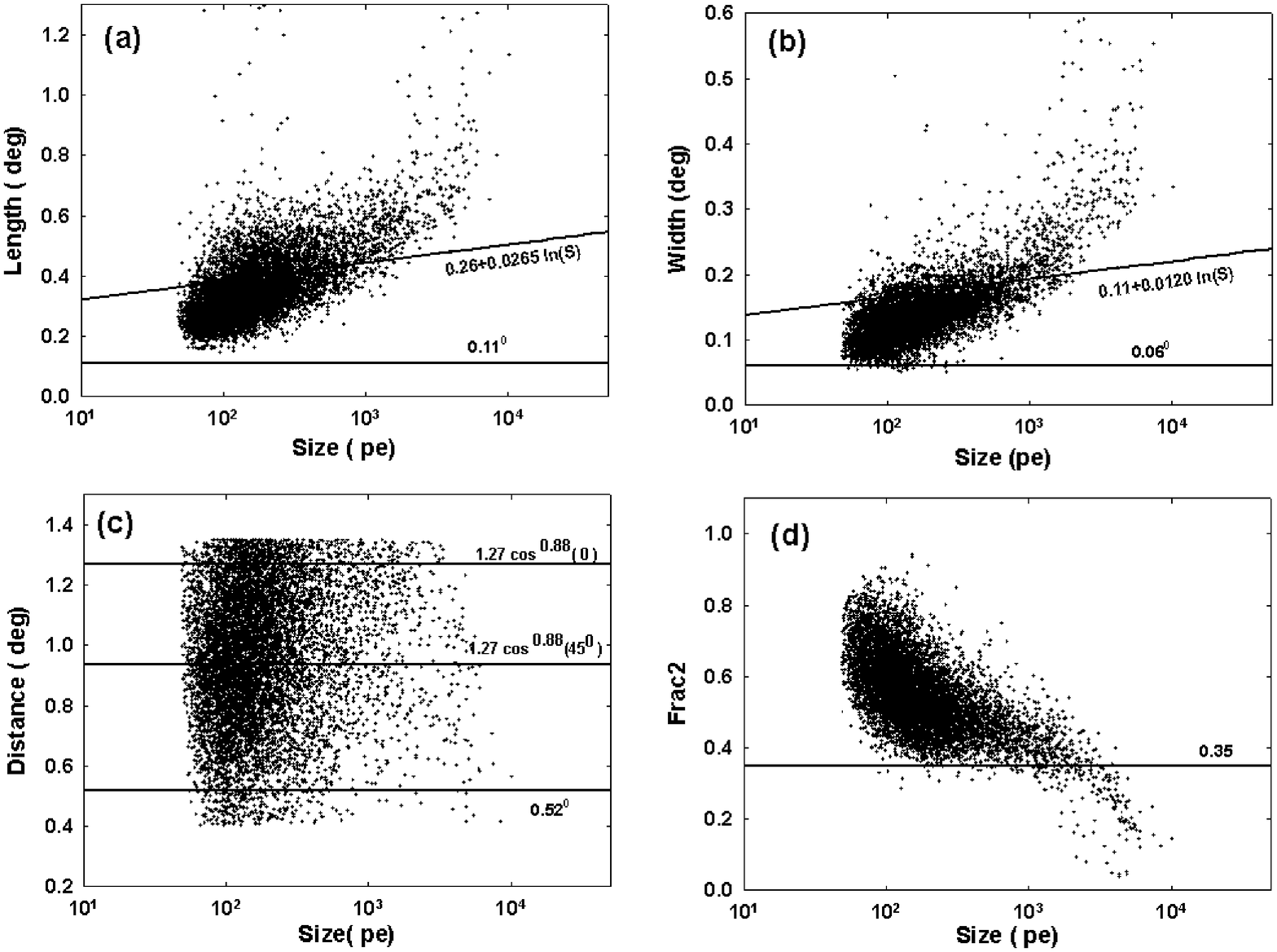}
\caption{\label{fig ---}  Scatter plots of  (a) LENGTH (b) WIDTH (c) DISTANCE and  (d) FRAC2  as a function of   SIZE
which   have been characterized  as   $\gamma$-ray like  by the  ANN  and   have  $\alpha\leq18^\circ$. 
The  Dynamic Supercuts  boundaries  are also  shown in  the figure as full lines.}  
\end{figure} 
For comparison,  the  Dynamic Supercuts  boundaries  are also  shown in  the figure as full lines. 
It is  quite  evident from  the figure  that the ANN  method  in not  just selecting   the same population of events as the  
 Dynamic Supercuts but  the ANN  is also   sensitive  to selecting events  which  lie outside  the   strict  Dynamic Supercuts  boundaries.  An alternative way  to assess the residual  population of  events  selected by  ANN is to perform a logical NOT selection between  the ANN and the Dynamic  Supercuts methods. On  performing  this  selection  the  number of $\gamma$-ray events  are  found out  to be   $\sim$(453$\pm$74) with a statistical significance of $\sim$6.27$\sigma$  which   again  suggests   that the ANN method  is  more useful  than the Dynamic Supercuts methods  while determining  the  energy spectrum of $\gamma$-ray source.
On  performing    a logical AND  selection  between  the ANN and the Dynamic  Supercuts methods  the  number of $\gamma$-ray events  yielded  are  $\sim$(655$\pm$71) corresponding  to   a statistical significance of $\sim$9.50$\sigma$.
\par
In order to understand  the  performance of ANN  for   $\gamma$-rays  at higher  energies (i.e,  the events which eventually  contribute  to the last  3 energy bins  of  Fig.8)  Fig.11  displays the scatter plot  of   $\sim$606 events which  have  been characterized  as   $\gamma$-ray like  by the  ANN    and  which have  their   $\alpha\leq18^\circ$. 
\begin{figure}[h]
\centering
\includegraphics*[width=1.0\textwidth,angle=0,clip]{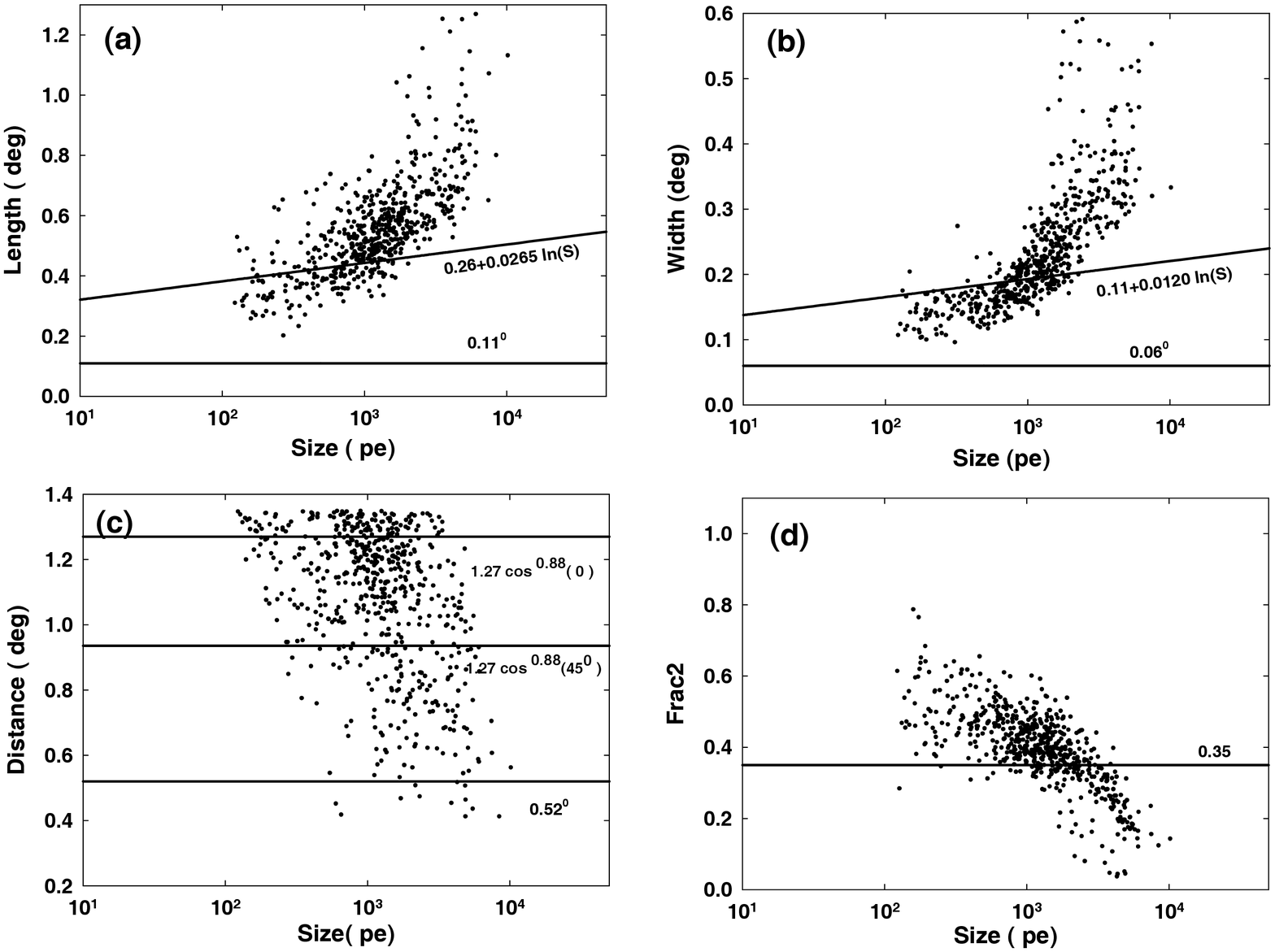}
\caption{\label{fig ---}  Scatter plots of  (a) LENGTH (b) WIDTH (c) DISTANCE and  (d) FRAC2  as a function of   SIZE
which    have   been characterized as  $\gamma$-ray like   by the  ANN. Apart from  having  $\alpha\leq18^\circ$ 
these events  also have   energy  above $\sim$9 TeV.  The  Dynamic Supercuts  boundaries  are also  shown in  the figure as full lines.}  
\end{figure} 
In other words the data presented  in  this  figure  represents  a subsample of the data used in Fig.10 with  an additional  condition 
that  the  $\gamma$-ray like  events  should  have   energies  above $\sim$9 TeV.  The capability of  the  ANN  in  selecting events  which  lie outside  the   strict  Dynamic Supercuts  boundaries  is   again  evident  from  the  figure. For example,  presence of relatively large number of event outside  the  LENGTH   cut boundary (Fig.11a)    clearly demonstrates  that  the   efficiency  of  Dynamic Supercuts  in retaining  $\gamma$-rays   is  biased  towards  lower   energies. 
It is important  to point  here  that  there are  background  cosmic-ray events  also  present  in  Fig.10 and Fig.11  which  are   classified as  $\gamma$-ray like events  by   the  event selection methodology.   Since   subtraction  of the background events  (estimated from         $27^\circ\leq \alpha \leq81^\circ$  region),   from   the $\gamma$-ray domain   (defined  as $\alpha\leq18^\circ$),  will  cancel out  these events  (in statistical sense) and  it  does not  matter how  the energy estimate  for  background event was  obtained.  
\par
Since  differences in the observed energy spectrum  of several  active galactic nuclei,  especially at higher energies, can   be used to study  absorption effects at the source or in the intergalactic medium due to interaction of $\gamma$-rays with the extragalactic 
background photons [73, 74],  unarguably,   efficient   retention  of   high energy  $\gamma$-ray  events is always preferable.   Superior  performance  of the   ANN   at  higher  energies  can  thus  play an important role   in   the  
understanding  the absorption  effects at the source or in the intergalactic medium.   
\par
It  is worth  mentioning  here that once satisfactory training of the ANN is  achieved, the corresponding ANN generated weight-file can  be  easily used  by  an  appropriate  subroutine  of the  main data analysis program  for selecting  $\gamma$-ray like events.  Use of  a
dedicated   ANN software package  is  thus  necessary  only  during   the  training  of  the  ANN and is not needed there after. 
Also, compared  to the conventional  $\gamma$/hadron separation methods, the  ANN-based   procedure   also  offers  advantages  like    applicability   over  a  wider  zenith angle  range  and  implementation  ease.        
\section{Conclusions}
Atmospheric   Cherenkov  imaging   telescopes, especially Monoscopic systems, have to cope  up with  a  deluge of cosmic-ray background events  and   the   capability to suppress these against the  genuine  $\gamma$-rays  is one of the main  challenges  which  limits the sensitivity of  these telescopes.  The main purpose of this paper is to study the $\gamma$ / hadron segregation potential of various  ANN algorithms for the TACTIC  telescope, by applying them to the Monte Carlo simulated  and the observation data  on the Crab Nebula.   The results of our study   indicate that   the  performance  of   Levenberg-Marquardt  based     ANN  algorithm  is somewhat  superior to   the  Dynamic Supercuts procedure especially beyond $\gamma$-ray energies of $\geq$ 9 TeV. 
Since for real world problems it is not an easy task to  identify the most  suitable  ANN algorithm by just having a look at the  problem, our results  suggest  that  while  investigating the comparative performance  of  other ANN algorithm,  the  Levenberg-Marquardt algorithm  deserves  a  serious   consideration.
The main  advantage  of  using   the  ANN methodology for $\gamma$/ hadron segregation  work  is that   it  is  more  efficient  in  retaining  higher energy $\gamma$-ray events    and   this  has   allowed  us to extend  the TACTIC observed energy  spectrum  of the Crab Nebula  up to an energy  of $\sim$24TeV.  
Reasonably good  matching of the  Crab Nebula  spectrum as  measured  by the TACTIC   telescope with that obtained  by the other  groups  reassures  that the   
ANN-based  $\gamma$/hadron segregation  method   and  also  the  procedure for obtaining the energy spectrum  of a $\gamma$-ray  source  are  quite  reliable. 
   
\section{Acknowledgements}
The authors would like to convey their gratitude to all the concerned colleagues of the  Astrophysical Sciences  Division  for their contributions towards the instrumentation, observation and analysis  aspects of the TACTIC telescope.  The authors would also like to thank the anonymous referees for their valuable comments which have helped us to improve the quality of the paper. 

\end{document}